\documentclass[english]{article}
\usepackage{geometry}
\usepackage{float}
\pdfoutput=1
\usepackage{mathrsfs}
\usepackage{amsmath}
\usepackage{amssymb}
\usepackage{graphicx}
\usepackage{amsfonts}
\usepackage{bm}
\usepackage{subfigure}
\usepackage{amsthm}
\usepackage{algorithm}
\usepackage{algorithmic}
\usepackage{etoolbox}
\usepackage{booktabs}
\usepackage{soul}
\usepackage{epsfig}
\makeatletter

\@ifundefined{definecolor}{\@ifundefined{definecolor}
 {\@ifundefined{definecolor}
 {\usepackage{color}}{}
}{}
}{}

\usepackage{subfig}\usepackage[all]{xy}

\newcounter{hypA}

\usepackage{babel}\date{}

\begin{document}

\begin{center}

{\Large \textbf{Estimating Option Prices using Multilevel Particle Filters}}

\bigskip

BY PRINCE PEPRAH OSEI $^{1}$ \&  AJAY JASRA $^{1}$ 

$^{1}${\footnotesize Department of Statistics \& Applied Probability,
	National University of Singapore, Singapore, 117546, SG.}\\
{\footnotesize E-Mail:\,}\texttt{\emph{\footnotesize op.peprah@u.nus.edu\footnote[2]{Main Author}, staja@nus.edu.sg}}
\end{center}

\begin{abstract}
Option valuation problems are often solved using standard Monte Carlo (MC) methods.  These techniques can often be enhanced using several strategies especially when one discretizes the dynamics
of the underlying asset, of which we assume follows a diffusion process. We consider the combination of two methodologies in this direction. The first is the well-known multilevel Monte Carlo (MLMC) method \cite{giles2008mlmc}, which is known to reduce the computational effort to achieve a given level of mean square error (MSE) relative to MC in some cases.  Sequential Monte Carlo (SMC) (or the particle filter (PF)) methods (e.g.~\cite{doucet_tutorial}) have also been shown to be beneficial in many option pricing problems potentially reducing variances by large magnitudes (relative to MC) \cite{JayOption,Sen_option}.  We propose a multilevel Particle Filter (MLPF) as an alternative approach to price options.   The computational savings obtained in using MLPF over PF for pricing both vanilla and exotic options is demonstrated via numerical simulations.
\end{abstract}

\section{Introduction} 
An option is a financial derivative which gives the option holder the right to buy or sell a specified amount of an underlying asset at a fixed price on or before the expiration date of the option.  To price an option is to evaluate the integral of its expected payoff under a risk-neutral probability measure, if such a measure exists (which is assumed).  In many practical applications, the underlying financial asset, which we shall assume throughout in this article, can be modelled by a diffusion process, or a pair of correlated diffusion processes (for instance stochastic volatility models).  The value of the financial option is the expectation of the underlying along a (discrete) path under the risk neutral meausure associated to the process just described.

As the value of the option is seldom available, one often resorts to numerical methods to approximate it.  Monte Carlo methods for pricing options dates back  at least to \cite{Boyle}.  These methods and its variants: quasi Monte Carlo (QMC), stratified sampling, control variate, antithetic variates and so on have been used extensively in the financial engineering literature; see for example \cite{glasserman_mcfinance} for a complete description of these techniques.  The main advantage of MC methods for pricing options compared to other numerical methods is its ability to deal with high-dimensional integrals. It is this methodology that is focussed upon
for the duration of the article.

In some recent works the standard MC method has been improved upon, especially when the diffusion process must be time-discretized.  In this latter scenario, the method of MLMC has (for some pay-offs and diffusions) been shown to provide the same overall MSE as an MC method, but with less computational effort; see for instance \cite{Giles_mlmc} and the references therein. We briefly note that the MLMC method works by considering a hierarchy of time-discretizations and a simple collapsing sum representation of the expectation w.r.t.~the most precise time-discretized diffusion process. For each summand, a difference of expectations under successively fine discretizations, the joint law of the discretized processes are coupled and sampled. This coupling, if sufficiently efficient can mean that the MLMC method achieves the afore-mentioned savings.  In addition to this, several works beginning with \cite{JayOption} and more recently in \cite{Sen_option,shev_op} have shown that standard MC and importance sampling (IS) can be enhanced by using SMC or PF methodology. This is an algorithm that can approximate expectations w.r.t.~a sequence of probability distributions by sampling a collection of $N$ particles (samples) in parallel, sequentially in time using sampling and resampling operations; see \cite{doucet_tutorial} for an introduction.  The main improvement of SMC over IS for option pricing is that as the number of points of the path of the diffusion process grows, call it $n$ (and under several mathematical assumptions) the relative variance of the SMC estimate is $\mathcal{O}(n/N)$ 
whereas for IS it can be $\mathcal{O}(\kappa^n/N)$ for some $\kappa>1$ (see e.g.~\cite{cdg:11}).
In this article, we show how using the works of \cite{Jay_mlpf,Jay_mlnormconst}, called the MLPF (see also \cite{Sen_coupling}), MLMC and SMC can be combined to help to improve upon the estimation of options.

The main contribution of this article is two fold:
\begin{enumerate}
	\item[1.] to show how the MLPF technique can be used in pricing European type exotic options.
	\item[2.] to illustrate by numerical examples the gains obtained when such methods are used to estimate option prices over vanilla particle filter.
\end{enumerate}

In terms of $1$, we aim to demonstrate that Multilevel Particle Filter (MLPF) framework can be used to estimate price of exotic options.  We compare  the benefits of using MLPF compared with standard particle filter (introduced in \cite{JayOption}) in estimating both basic and path dependent options.   With respect to $2$, numerical simulations of these two algorithms are introduced when the stochastic volatility is used, in particular a Langevin stochastic differential equation (SDE).  
*We are not intended to show that these methods presented here are competitors to the existing ones but demonstrate that it enhances the existing methods in the literature.

The rest of the article is structured as follows.  Section \ref{ch3:sec1} introduces the option pricing problem and two path dependent options, barrier options and target accrual redemption notes (TARNs).  It also reviews identities for approximation for evaluating the expected value of the functional of the underyling processes.  Section \ref{ch3:sec3} briefly reviews the multilevel particle filter technique relevant in this context.  Section \ref{ch3:sec4} numerically illustrates our methods.  The article concludes in Section \ref{ch3:sec5}, and presents further extensions for future research.

In what is to follow in the rest of the paper, the following notations will be adopted.  For any vector $x_{1:n}$ will denote $\left(x_1,\dots,x_n\right)$.  Expectations are written generically as $\mathbb{E}$ and subscript is added, if it is required to denote dependence upon a measure/point.  $\mathbb{R}^d$ denote the $d$-dimensional Euclidean space.  For $k\in \mathbb{N}$, $\mathbb{T}_k=\left\lbrace1,\dots,k\right\rbrace$.

\section{The Model, Options and Strategy}\label{ch3:sec1}

\subsection{The Price Process}

Let $\left\lbrace S_t\right\rbrace_{t\in\left[0,T\right]}$ be the price process, $S_t\in\mathbb{R}^{+}$ of an underlying financial asset.  We assume that it follows a diffusion:
\begin{eqnarray}\label{ch3:eq1}
\mathrm{d}S_t &= & \alpha\left( S_t\right)\mathrm{d}t+\beta\left( S_t,V_t\right)\mathrm{d}W_t \nonumber \\
\mathrm{d}V_t &= & \gamma\left( V_t\right)\mathrm{d}t+\nu\left( V_t\right)\mathrm{d}B_t 
\end{eqnarray}
where $\alpha\thinspace\colon\thinspace\mathbb{R}^{+}\rightarrow \mathbb{R}$, $\beta\thinspace\colon\thinspace(\mathbb{R}^{+})^2\rightarrow \mathbb{R}^{+}$ 
$\gamma\thinspace\colon\thinspace\mathbb{R}\rightarrow \mathbb{R}$, $\nu\thinspace\colon\thinspace\mathbb{R}\rightarrow \mathbb{R}^{+}$,
with $\{V_t\}_{t\in\left[0,T\right]}$ is the volatility (or log-volatility)
and $\left\lbrace W_t\right\rbrace_{t\in\left[0,T\right]}$, $\left\lbrace B_t\right\rbrace_{t\in\left[0,T\right]}$ are independent standard Brownian motion. Throughout $V_0,S_0$ is assumed known.  In addition, the functions $\alpha,\beta,\gamma$ and $\nu$ are assumed known (see, e.g.~\cite{Barndorff_OUbased,JayOption,shev_op}).  In some examples there will not be any volatility process.

We are interested in computing options of the form
\begin{align*}
\mathbb{E}\left[g\left(S_{t_1:t_k}\right)\right]
\end{align*}
for some $k\geq 1$ given and $g:(\mathbb{R}^{+})^k\rightarrow\mathbb{R}^{+}$ and the expectation is typically w.r.t.~the time discretized process (e.g.~Euler discretization).

\subsection{Barrier Options}
Barrier options are derivatives for which the payoff may be zero dependent on the path of the underlying asset $\left\lbrace S_t\right\rbrace_{t\in I}$, $I\subset\left[0,T\right]$, breaching a barrier.  There are two broad types of barrier options:
\begin{enumerate}
	\item[1.] knock-in: the option pays zero unless a function of the underlying asset values breaches prespecified barriers (option springs into existence) and 
	\item[2.] knock-out: the option pays zero if a function of the underlying asset values breaches prespecified barriers (option is extinguished).
\end{enumerate}
Compared to basic options, barrier options are cheaper because it may expire worthless if knocked out (or knocked in) in the same condition in which the vanilla option would have paid off.

Barrier options are hard to price using standard MC methods due to most particles leading to a zero payoff and this gives inaccurate estimates of the option.  As noted in \cite{Glasserman_barrieroptions} many paths may lead to zero Monte Carlo and simple remedy suggested is to use conditional distribution gvien one-step survival.

\subsubsection{Barrier Options with constant Volatility}
In this paper, we will concentrate upon European style options
\begin{align*}
\mathbb{E}_{S_0}\left[g\left(\left\lbrace S_t\right\rbrace_{t\in I}\right)\right]
\end{align*}
with
\begin{align*}
g\left(\left\lbrace S_t\right\rbrace_{t\in I}\right)&=\mathbb{I}_{B}\left(\left\lbrace S_t\right\rbrace_{t\in I}\right)e^{-rT}\left(S_T-K\right)_{+}
\end{align*}
a barrier call option, with strike $K>0$, interest rate $r>0$ and $B$ the barrier set.  Consider, for instance in a simpler case, a constant volatility and a discretely monitored   knock-out barrier option, $I=\left\lbrace t_1,\dots,t_k\colon 0<t_1<\dots<t_k=T\right\rbrace$, $t_0=0$, barrier set $B=\bigotimes_{i=1}^{k}\left[L_{t_i},U_{t_i}\right]$, the option price is:
\begin{align}\label{ch3:eq2}
\int e^{-rT}\left(s_{t_k}-K\right)_{+}\prod_{i=1}^{k}\left\lbrace\mathbb{I}_{\left[L_{t_i},U_{t_i}\right]}\left(s_{t_i}\right)p\left(s_{t_i}\mid s_{t_{i-1}}\right)\right\rbrace\mathrm{d}\left(s_{t_{1:k}}\right),
\end{align}
where we assumed that the unknown transition densities can be written with respect to a dominating measure abusively denoted here as $\mathrm{d}\left(s_{t_{1:k}}\right)$.  The estimation of the barrier option \eqref{ch3:eq2} is non-trivial in Monte Carlo integration.  For instance, if we assume all the parameter functions and transition densities are known and the Euler discretization adopted, it is still the case that many paths may yield to a zero Monte Carlo estimate before the terminal time.  We remark here that, a more complicated model of the volatility process can be adopted; see for example \cite{Barndorff_OUbased,JayLevy}.  We will consider in addition to the constant volatility model the case where the volatility process is a stochastic differential equation, in particular a Langevin SDE.

\subsection{Target and Accrual Redemption Options}
These options are very popular in the FX trading market.  It provides  the holder a capped sum of payments over a period with the possibility of premature termination (knock-out) due to target cap pre-specified on the accumulated payments.  A specified amount of payment is made on coupon dates (referred to as fixing dates or cash flow dates) until the cap target is violated.   One typical example is the target accrual redemption note.  The note value on a coupon date depends on the spot value of the underlying asset and the accumulated payment amount up to the coupon date.

For illustration purposes, we consider the typical TARNs introduced in \cite{Sen_option}.  Consider a sequence of cash flow dates $\left\lbrace T_n\right\rbrace_{n\in\mathbb{T}_k}$, where $T_k=T$ is the note's maturity date, and a real-valued function $f\thinspace\colon\mathbb{R}\rightarrow\mathbb{R}$.  The gain and losses processes are given by 
\begin{align*}
G_p&=\sum_{i=1}^{p}f^{+}\left(S_{T_i}\right),\quad L_p=\sum_{i=1}^{p}f^{-}\left(S_{T_i}\right),
\end{align*}
where $f^+, f^{-}$ are positive and negative parts of $f$ and $G_p, L_p$ are positive and negative cash flows respectively.  
Set
$$
\tau^{(L)} = \inf\{k\geq 1:L_k\geq \Gamma_L\}\quad \tau^{(G)} = \inf\{k\geq 1:G_k\geq \Gamma_G\}
$$
for two given constants $\Gamma_L,\Gamma_G$. Set
$$
\tau = \min\{\tau^{(L)},\tau^{(G)},k\}.
$$
The value of the TARN is  the expected value of the overall cash flow
\begin{align*}
\mathbb{E}\left[\sum_{i=1}^{\tau}e^{-r\tau}f\left(S_{T_i}\right)\right],
\end{align*} 
where $r$ is the interest rate.  We have assumed that the interest rate is deterministic.  A more practical version can be considered where the rate is underlying stochastic state variable.  Additionally, the amount paid to the holder and the termination date is uncertain. Note that this can be written in the form of interest. Let
$$
A_i = \{S_{1:i}\in(\mathbb{R}^+)^i:G_i<\Gamma_G\cap L_i<\Gamma_L\}
$$
then the value of the TARN is
$$
\mathbb{E}\left[\sum_{i=1}^{k}e^{-r\tau}\mathbb{I}_{A_i}(S_{T_1:T_i})f\left(S_{T_i}\right)\right].
$$

Most TARN's price estimation are solved using standard MC techniques.  In the situation where the function $f$ is discontinous, the MC estimates are unreliable and thus SMC techniques offers the best alternative to the problem.  In particular, we consider both the standard particle filter and multilevel particle filter when the volatility is constant and modelled by a stochastic differential equation.

\subsection{Identities for Approximation}\label{ch3:sec2}
We will suppose that the process \eqref{ch3:eq1} is suitably exotic such that 
\begin{enumerate}
	\item[1.] One cannot compute $\mathbb{E}\left[g\left(S_{t_1:t_k}\right)\right]$
	\item[2.] One cannot sample from the law of $S_{t_1:t_k}$ exactly, that is, without discretization error.
\end{enumerate}
It is suppose that one will discretize (e.g.~Euler or Milstein) and call the time discretization $h_L$, with the exact solution of $\left(\ref{ch3:eq1}\right)$ returned when $h_L=0$.  We then take expectations with respect to a law with the following finite dimensional law:
\begin{align}\label{ch3:eq3}
\prod_{i=1}^{k} Q^L\left((v_{i-1},s_{i-1}),(v_i,s_i)\right),
\end{align}
where $ Q^L$ is the transition kernel induced by the time discretization.  Note that we are simply denoting $V_1,S_1,V_2,S_2,\dots$ as the random variables associated to the time discretization: the actual time in $\left[0,T\right]$ is suppressed from the notation.  So we are reduced to computing
\begin{align*}
\mathbb{E}_L\left[g(S_{1:k})\right],
\end{align*}
where the expectation is with respect to the law associated to \eqref{ch3:eq3}.  As noted in \cite{Jay_smcdiff} even if one can sample from the law of $S_{1:k}$ exactly, using standard Monte Carlo can induce a substantial variance.  This is also true when discretizing the time parameter.

\subsubsection{Importance Sampling}
We suggest the following, close to optimal (in some sense), importance sampling procedure as used in \cite{JayOption,Jay_mlpf}.  Let $\tilde{g}\thinspace\colon\thinspace(\mathbb{R}^{+})^k\rightarrow\mathbb{R}^{+}$ be a function `related' to $g$ (which could be $g$ itself).  Then consider the target density
\begin{align*}
\pi_L\left(s_{1:k},v_{1:k}\right)\propto\kappa_L\left(s_{1:k},v_{1:k}\right)&=\tilde{g}\left(s_{1:k}\right)\prod_{i=1}^{k} Q^L\left((v_{i-1},s_{i-1}),(v_i,s_i)\right).
\end{align*}

We note that:
\begin{align*}
\mathbb{E}_L\left[g\left(S_{m_L}\right)\right]&=Z_{\pi_L}\mathbb{E}_L\left[\frac{g\left(S_{1:k}\right)}{\tilde{g}\left(S_{1:k}\right)}\right]
\end{align*}
where $Z_L=\int_{\left(\mathbb{R}^{+}\right)^{M_L}}\kappa_L\left(s_{1:M_L}\right)\mathrm{d}s_{1:M_L}$. It is this identity that we will
try to approximate efficiently. In \cite{Jay_smcdiff,JayOption}, it is discussed why such an approach can be useful. For instance, as used in \cite{Glasserman_barrieroptions},
in the context of barrier options a change of measure is useful to ensure that samples from the change of measure will stay in a region of importance,
i.e.~yield (relatively) low variance Monte Carlo estimates. Note that from a minimum variance perspective (i.e.~for importance sampling), the optimal choice is $\tilde{g}\left(s_{1:k}\right)=g\left(s_{1:k}\right)$,
but this may not always work well; see e.g.~\cite{JayOption} for some discussion.

\subsubsection{Multilevel Identity}\label{sec:ml_id}
Consider a hierarchy of discretizations $0<h_L<\dots<h_1<+\infty$ with obvious extension of $\pi_L, \kappa_L, Z_L$.  Then one has that
\begin{align*}
\mathbb{E}_L\left[g\left(S_{1:k}\right)\right]&=\sum_{l=1}^{L}\left(\mathbb{E}_l\left[g\left(S_{1:k}\right)\right]-\mathbb{E}_{l-1}\left[g\left(S_{1:k}\right)\right]\right)\\
&=\sum_{l=1}^{L}\Bigg\{
Z_{\pi_l}\mathbb{E}_l\left[\frac{g\left(S_{1:k}\right)}{\tilde{g}\left(S_{1:k}\right)}\right] - 
Z_{\pi_{l-1}}\mathbb{E}_{l-1}\left[\frac{g\left(S_{1:k}\right)}{\tilde{g}\left(S_{1:k}\right)}\right]
\Bigg\},
\end{align*}
where we use the convention that $Z_{\pi_{0}}\mathbb{E}_{0}\left[\frac{g\left(S_{1:k}\right)}{\tilde{g}\left(S_{1:k}\right)}\right]=0$.  \cite{Jay_mlnormconst} show how the right hand side of the final line can be approximated using the multilevel particle filter \cite{Jay_mlpf}.  They also show the theoretical benefits relative to using the procedure in the previous section. This is explicitly in the case of a pair of possibly correlated diffusion processes and under regularity conditions.  It shown that for a specific choice of $L$ and number of samples at each level this MLPF has the same MSE as
a PF but with less computational effort. We note that in the MLPF approach to be discussed in the next Section, these results also apply under the conditions in \cite{Jay_mlpf,Jay_mlnormconst}. 

\section{Pricing Options using Multilevel Particle Filters}\label{ch3:sec3}
In this section, we briefly introduce the multilevel particle filter in particular context of estimating option prices.  We begin by briefly explaining the standard particle filter and then its extension to the multilevel particle filter framework.  We will suppose that for any $1\leq n \leq k$ one has a function $\tilde{g}_n:(\mathbb{R}^+)^n\rightarrow\mathbb{R}^+$
associated to $\tilde{g}$. For instance, in the case of barrier options, we shall take
$$
\tilde{g}(s_{1:k}) = |s_k-K|^{\rho}\prod_{i=1}^k\mathbb{I}_{[L_i,U_i]}(s_i)
$$
for some fixed $\rho\in(0,1)$ so there is a natural extension
$$
\tilde{g}_n(s_{1:n}) = |s_n-K|^{\rho}\prod_{i=1}^n\mathbb{I}_{[L_i,U_i]}(s_i).
$$

\subsection{Estimating Option prices using the PF}
A standard PF is given in Algorithm \ref{algo:ch3:SPF} for a given $l\geq 1$; we use the notation
$x_n=(v_n,s_n)$. 
The resampling step is described in detail in e.g.~\cite{doucet_tutorial} and can be made adaptive, i.e.~only resampling
`when needed'.
 The unbiased estimator \cite{delm:04} of $\mathbb{E}_l[g(S_{1:k})]$ is
$$
\Big(\prod_{p=1}^{k-1} \frac{1}{N_l}\sum_{i=1}^{N_{l}} W_p^{l,i}\Big) \frac{1}{N_l}\sum_{i=1}^{N_{l}}W_k^{l,i}g(\check{S}_{1:k-1}^{l,i},S_k^{l,i}).
$$
Note that typically, this algorithm only works well if $\tilde{g}_n$ is of product form, or if $\tilde{g}_n$ depends only on $s_{n-u:n}$ for some small $u$ or 
$k$ is small; see \cite{delm:04} for the reasons why and a justification of the algorithm. One of these will be the case in all of our examples.

\begin{algorithm}[!ht]
	\caption{\textbf{A generic PF algorithm}}
	\label{algo:ch3:SPF}
	\begin{algorithmic} 
		\STATE
		\begin{itemize}
			\item 0. Set $n=1$; for each $i\in\mathbb{T}_{N^{l}}$ sample $X_1^{(l,i)}\sim Q^l((v_0,s_0),\cdot)$
			and set initial weights $W_{1}^{(l,i)}=\tilde{g}_1(s_1^{l,i})$.
			\item 1. Resample $X_1^{(l,1)},\dots,X_1^{(l,N_l)}$	 and write the resulting samples $\check{X}_1^{(l,1)},\dots,\check{X}_1^{(l,N_l)}$.
			\item 2. Set $n = n+1$; for each $i\in\mathbb{T}_{N^{l}}$ sample $X_n^{(l,i)}\sim Q^{l}\left(\check{x}_{n-1}^{l,i},\cdot\right)$ and compute weights $W_n^{(l,i)}=\tilde{g}_n(\check{s}_{1:n-1}^{l,i},s_n^{l,i})/\tilde{g}_{n-1}(\check{s}_{1:n-1}^{l,i})$.
			\item 3.  Resample $(\check{X}_{1:n-1}^{(l,1)},X_{n}^{(l,1)}),\dots,(\check{X}_{1:n-1}^{(l,N_l)},X_{n}^{(l,N_l)})$ and write the resulting samples 
$\check{X}_{1:n}^{(l,1)},\dots,\check{X}_{1:n}^{(l,N_l)}$. If $n=k$ stop, otherwise return to 2.
		\end{itemize}
	\end{algorithmic}
\end{algorithm}

\subsection{Estimating Option prices using the MLPF}

We now discuss how one may approximate the identity detailed in Section \ref{sec:ml_id}. In the case of the first summand, one can run the procedure detailed
in the previous section. So we focus on the approximation of a term of the form, for $2\leq l \leq L$:
$$
Z_{\pi_l}\mathbb{E}_l\left[\frac{g\left(S_{1:k}\right)}{\tilde{g}\left(S_{1:k}\right)}\right] - 
Z_{\pi_{l-1}}\mathbb{E}_{l-1}\left[\frac{g\left(S_{1:k}\right)}{\tilde{g}\left(S_{1:k}\right)}\right].
$$
We will use $L-1$ independent algorithms to approximate the above expression. Critical to our approach will
be the use of a coupling of the kernel $Q^l$. We will suppose that there is a $\check{Q}^{l,l-1}$ such that for
any $(x^l,x^{l-1})\in(\mathbb{R}_+\times\mathbb{R})^2$ one has for any set $A$
$$
\int_{A\times(\mathbb{R}_+\times\mathbb{R})}\check{Q}^{l,l-1}((x^l,x^{l-1}),(\tilde{x}^l,\tilde{x}^{l-1})) {d}(\tilde{x}^l,\tilde{x}^{l-1})
= \int_{A} Q^{l}(x^l,\tilde{x}^l) {d}\tilde{x}^l \quad\textrm{and}
$$
$$
\int_{(\mathbb{R}_+\times\mathbb{R})\times A}\check{Q}^{l,l-1}((x^l,x^{l-1}),(\tilde{x}^l,\tilde{x}^{l-1})) {d}(\tilde{x}^l,\tilde{x}^{l-1})
= \int_{A} Q^{l-1}(x^{l-1},\tilde{x}^{l-1}) {d}\tilde{x}^{l -1}.
$$
Such a coupling exists in our context, see the appendix. The algorithm is presented in Algorithm \ref{algo:ch3:ML}. Note that the coupled resampling procedure
is described in detail in \cite{Jay_mlpf,Jay_mlnormconst}. An unbiased estimate of $\mathbb{E}_l\left[g\left(S_{1:k}\right)\right]-\mathbb{E}_{l-1}\left[g\left(S_{1:k}\right)\right]$ is given by (see \cite{Jay_mlnormconst})
$$
\Big(\prod_{p=1}^{k-1} \frac{1}{N_l}\sum_{i=1}^{N_{l}} W_{p,l}^{l,i}\Big) \frac{1}{N_l}\sum_{i=1}^{N_{l}}W_{k,l}^{l,i}g(\check{S}_{1:k-1}^{l,i},S_k^{l,i}) - 
\Big(\prod_{p=1}^{k-1} \frac{1}{N_l}\sum_{i=1}^{N_{l}} W_{p,l}^{l-1,i}\Big) \frac{1}{N_l}\sum_{i=1}^{N_{l}}W_{k,l}^{l-1,i}g(\check{S}_{1:k-1}^{l-1,i},S_k^{l-1,i}).
$$

\begin{algorithm}[!ht]
	\caption{\textbf{A generic MLPF algorithm}}
	\label{algo:ch3:ML}
	\begin{algorithmic} 
		\STATE
		\begin{itemize}

\item 0. Set $n=1$; for each $i\in\mathbb{T}_{N^{l}}$ sample $(X_1^{(l,i)},X_1^{(l-1,i)})\sim \check{Q}^{l,l-1}((x_0,x_0),\cdot)$
			and set initial weights $W_{1,l}^{(l,i)}=\tilde{g}_1(s_1^{l.i}),W_{1,l}^{(l-1,i)}=\tilde{g}_1(s_1^{l-1.i})$.
			\item 1. Perform coupled resampling on  $(X_1^{(l,1)},X_1^{(l-1,1)}),\dots,(X_1^{(l,N_l)},X_1^{(l-1,N_l)})$	and write the resulting samples $
(\check{X}_1^{(l,1)},\check{X}_1^{(l-1,1)}),\dots,(\check{X}_1^{(l,N_l)},\check{X}_1^{(l-1,N_l)})$.
			\item 2. Set $n = n+1$; for each $i\in\mathbb{T}_{N^{l}}$ sample $(X_n^{(l,i)},X_n^{(l-1,i)})\sim \check{Q}^{l,l-1}\left((\check{x}_{n-1}^{l,i},\check{x}_{n-1}^{l-1,i}),\cdot\right)$ and compute weights $W_{n,l}^{(l,i)}=\tilde{g}_n(\check{s}_{1:n-1}^{l,i},s_n^{l,i})/\tilde{g}_{n-1}(\check{s}_{1:n-1}^{l,i})$, 
$W_{n,l}^{(l-1,i)}=\tilde{g}_n(\check{s}_{1:n-1}^{l-1,i},s_n^{l,i})/\tilde{g}_{n-1}(\check{s}_{1:n-1}^{l-1,i})$.
			\item 3. 
Perform coupled resampling on  $((\check{X}_{1:n-1}^{(l,1)},X_n^{l,1}),(\check{X}_{1:n-1}^{(l-1,1)},X_n^{l-1,1})),\dots,$
$((\check{X}_{1:n-1}^{(l,N_l)},X_n^{l,N_l}),(\check{X}_{1:n-1}^{(l-1,N_l)},\check{X}_n^{l-1,N_l}))$ and write the resulting samples 
$(\check{X}_{1:n}^{(l,1)},\check{X}_{1:n}^{(l-1,1)}),\dots,$ $
(\check{X}_{1:n}^{(l,N_l)},\check{X}_{1:n}^{(l-1,N_l)})$.  If $n=k$ stop, otherwise return to 2.
 		 
		\end{itemize}
	\end{algorithmic}
\end{algorithm}

\section{Simulations}\label{ch3:sec4}
In this section we demonstrate, numerically, the computational savings obtained in using the MLPF over the standard PF for option pricing.  In order to compare the mean-square error (MSE) estimate against the computational cost of Algorithms \ref{algo:ch3:SPF} and \ref{algo:ch3:ML}, we run each $50$ times.  We then look at MSE of the $50$ estimates and report the MSE versus computational cost.  In the sequel we consider the basic European call option, barrier options and target accrual redemption notes.  The approach in \cite{Jay_mlnormconst} to choose $L$ and the $N_{1:L}$ is adopted for the MLPF. A time discretization $h_l=2^{-l}$ is used. For both the PF and MLPF adaptive resampling
is used.

\subsection{Pricing Basic European Call Option}

We consider a standard European call option and the underlying is geometric Brownian motion and there is only constant volatility.
The functions $\tilde{g}_n$ are taken as $|S_n-K|^{\rho}$, where $K$ is the strike price; the choice is used to prevent the weights in the PF being zero (or infinite).
 We compare estimates from these two algorithms, PF and MLPF with exact option price from the Black-Scholes formula.  
Of course in this example, neither discretization nor MC are needed, but is just used as an example where the price is known.   The PF and ML are run at time discretization $h_5$ where $h_5=2^{-5}$ and consider time points $\mathbb{T}_{50}$ for illustration purposes.

\begin{figure}[!h]
	\centering
	\subfigure{{\includegraphics[height=7cm]{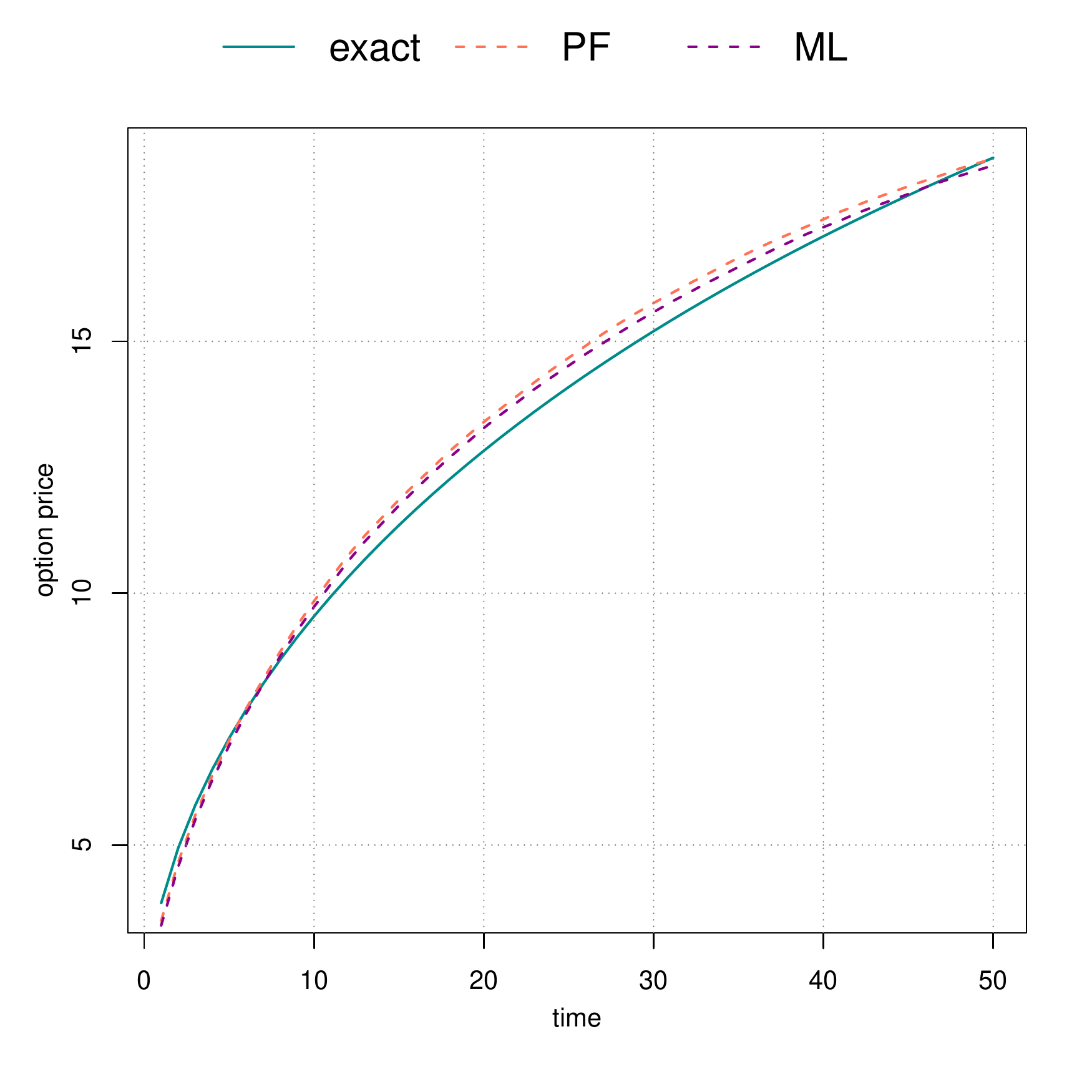}}}\,\,
	\subfigure{{\includegraphics[height=7cm]{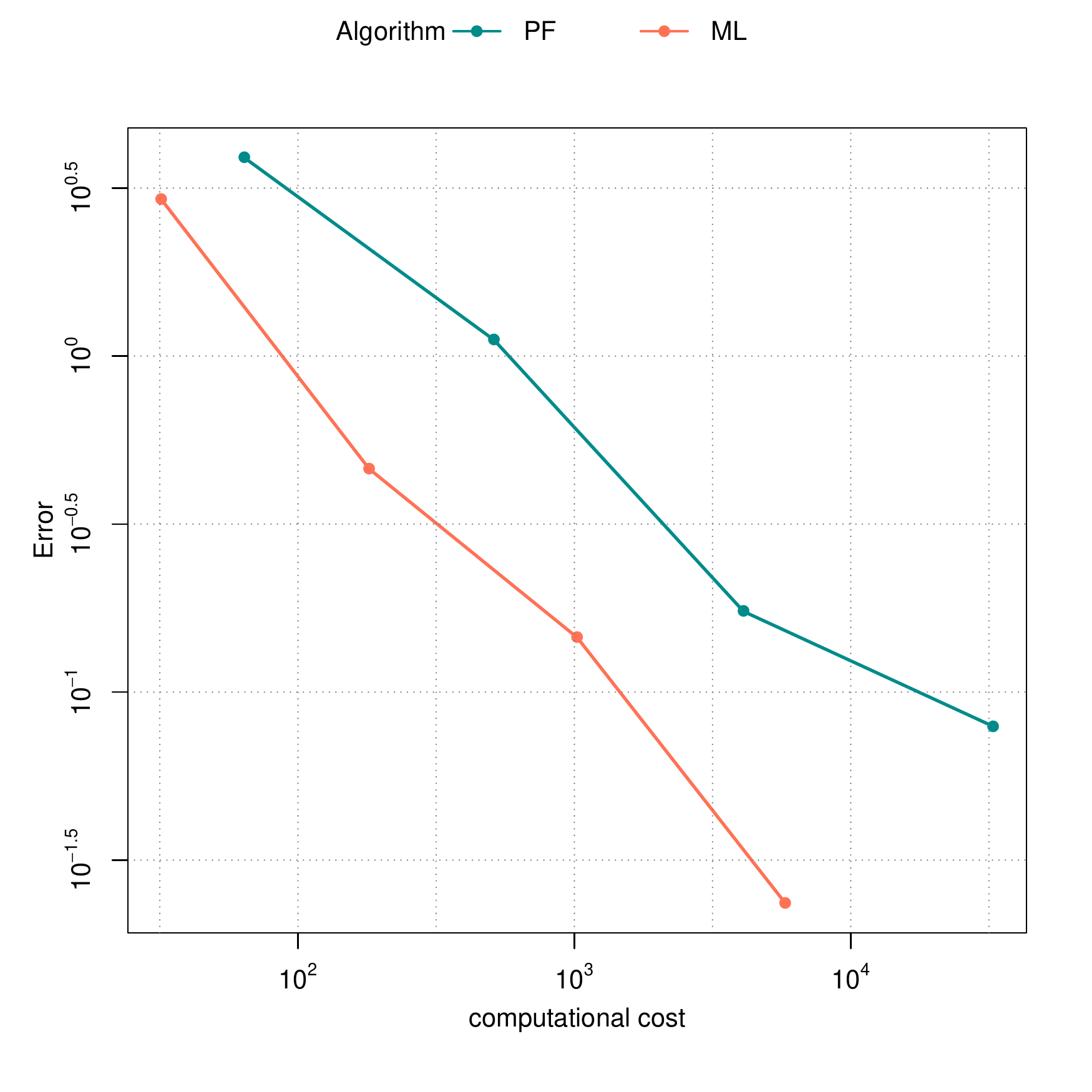}}}
	\caption{Estimation of basic European call option for time points $n=50$ at level $L=5$; in the right figure, the computational savings of using multilevel particle filter compared to standard particle filter.}
	\label{ch3:fig1}
\end{figure}

The results are shown in Figure \ref{ch3:fig1}.  We can observe that both algorithms estimate the price with good precision.  Considering the computational savings plot, we observe that the MLPF outperforms the standard PF in this basic option price estimate.  In the subsequent sections, more exotic options difficult to price will be considered.

\subsection{Barrier Options}
We consider a knock-out barrier option.  As noted in the earlier sections of this article, this is a path dependent option which standard MC gives inaccurate estimates since most of the particles give zero MC estimate.   The most common monitoring strategy adopted in the literature is to monitor the underlying assets after every $n$ units of time for a total of $k$ time periods.  We consider four different $n$ units of time, that is $n\in\left\lbrace50,75,100,200\right\rbrace$.

\subsubsection{Constant Volatility}

\begin{figure}[!ht]
	\centering
	\subfigure{{\includegraphics[height=7cm]{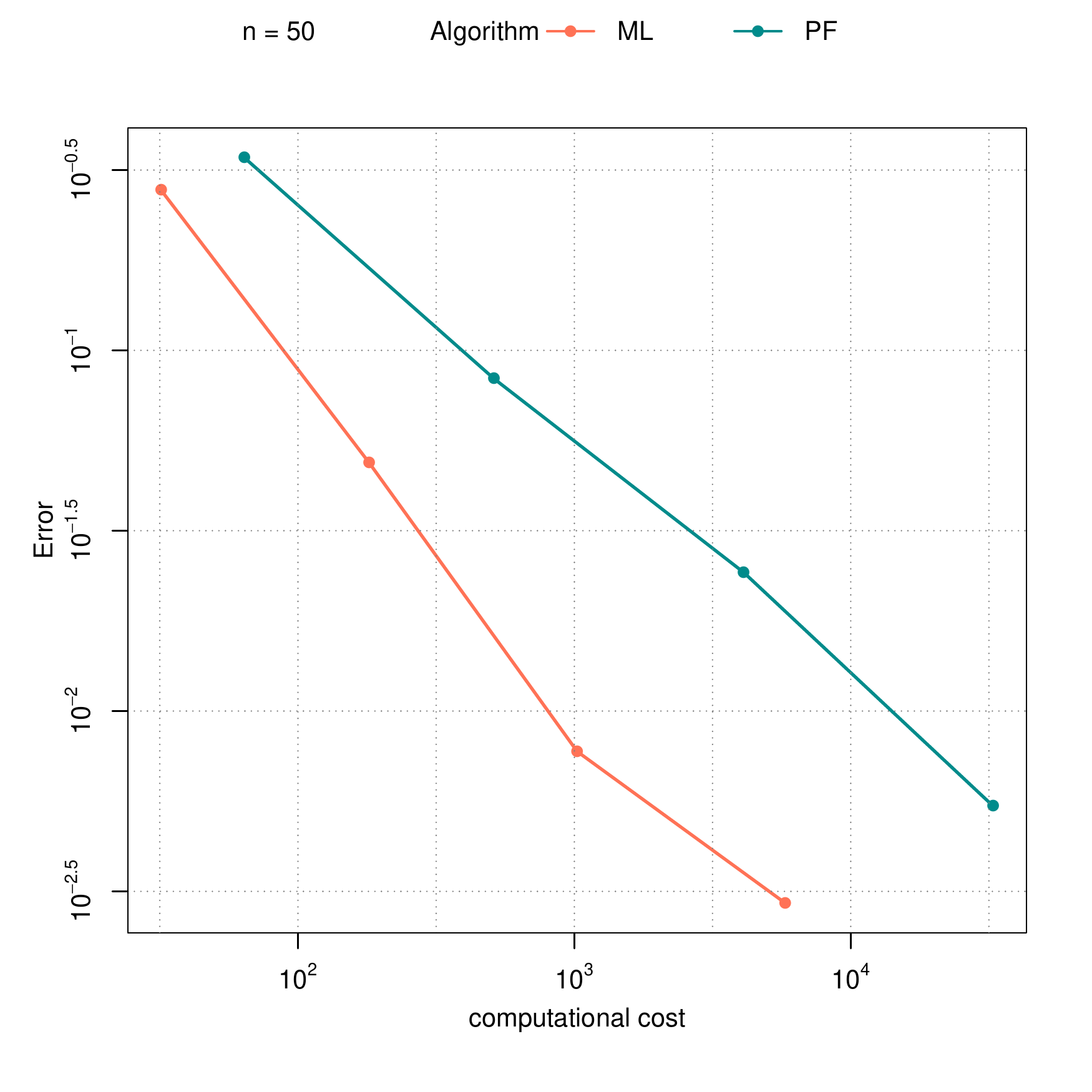}}}\,\,\,\,
	\subfigure{{\includegraphics[height=7cm]{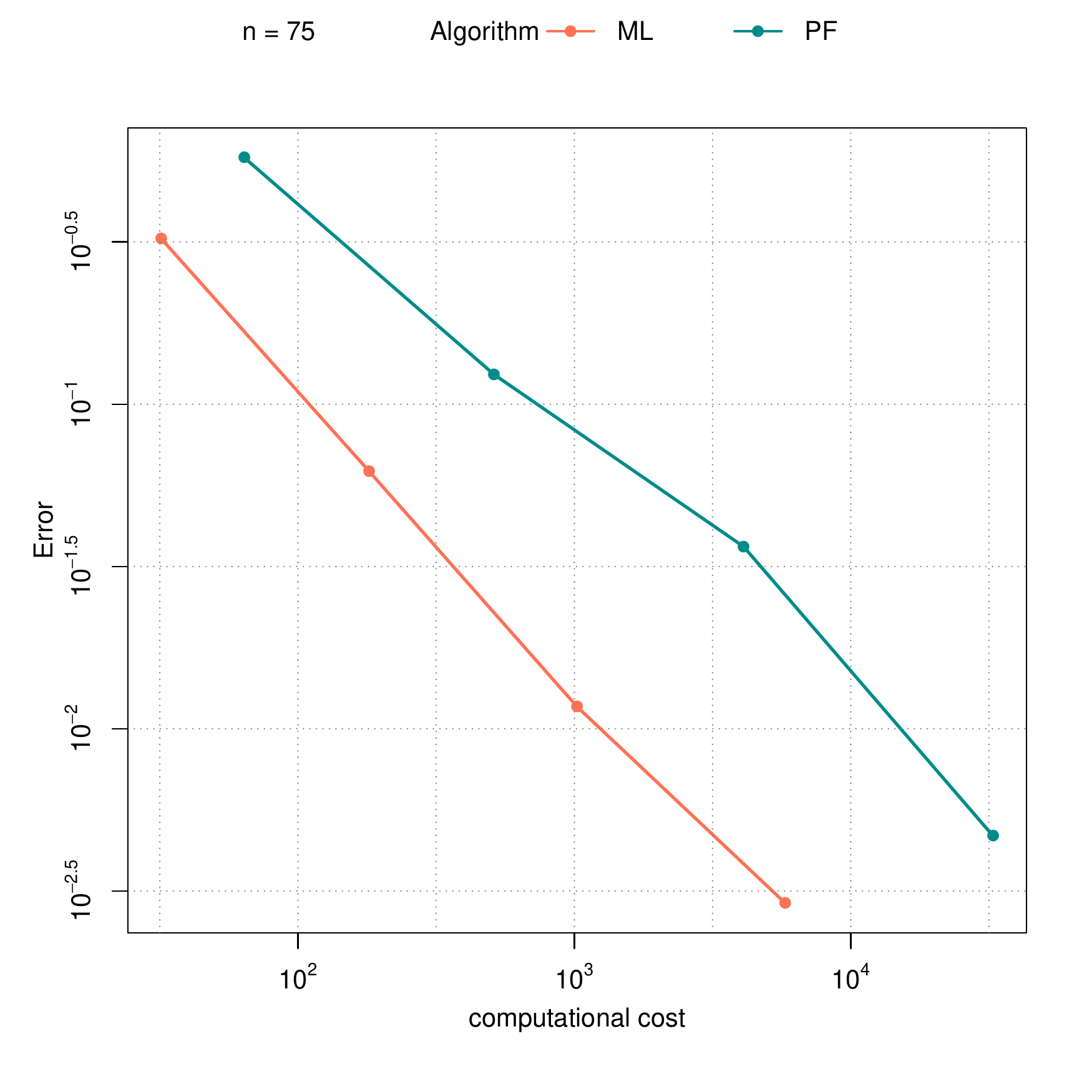}}}
	\subfigure{{\includegraphics[height=7cm]{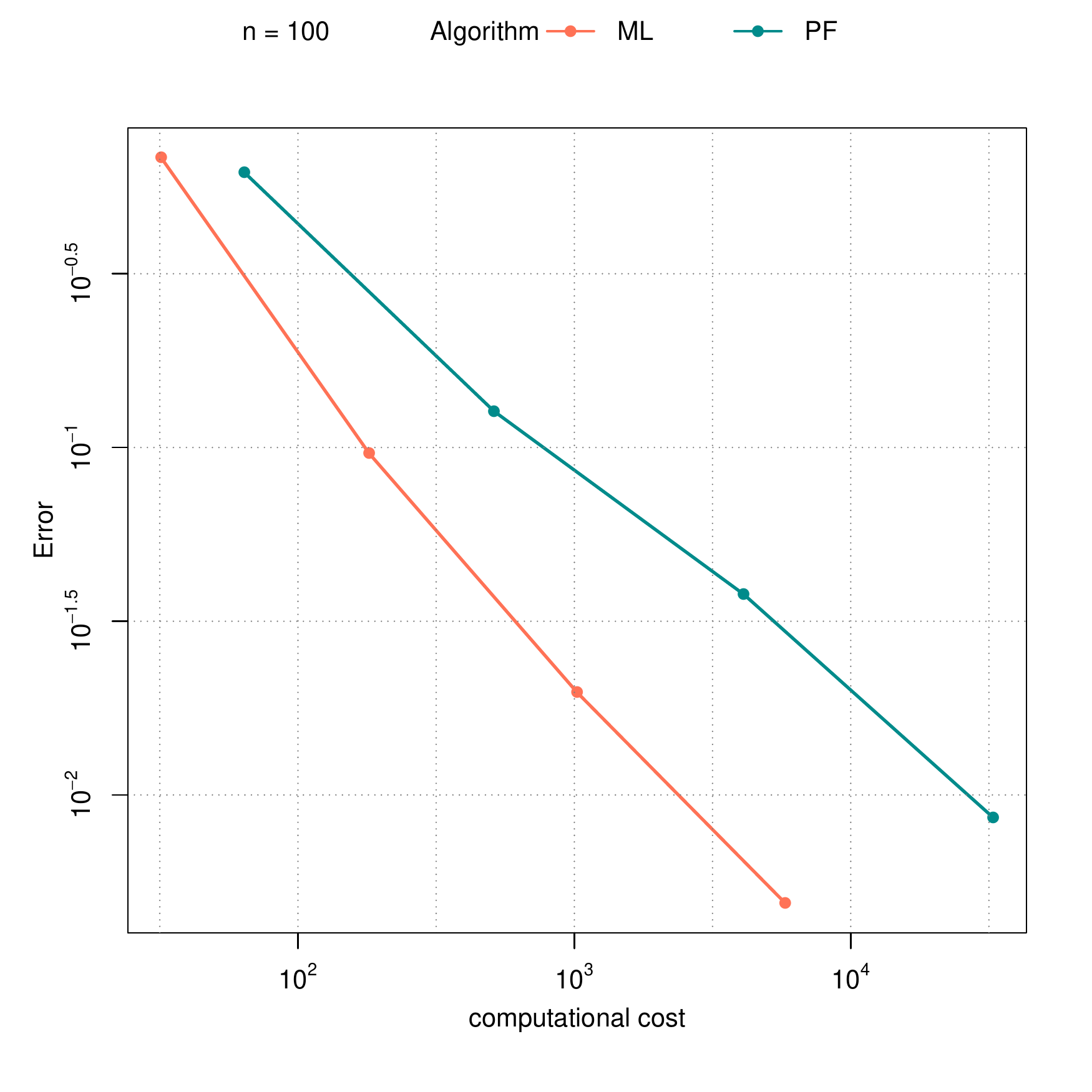}}}
	\subfigure{{\includegraphics[height=7cm]{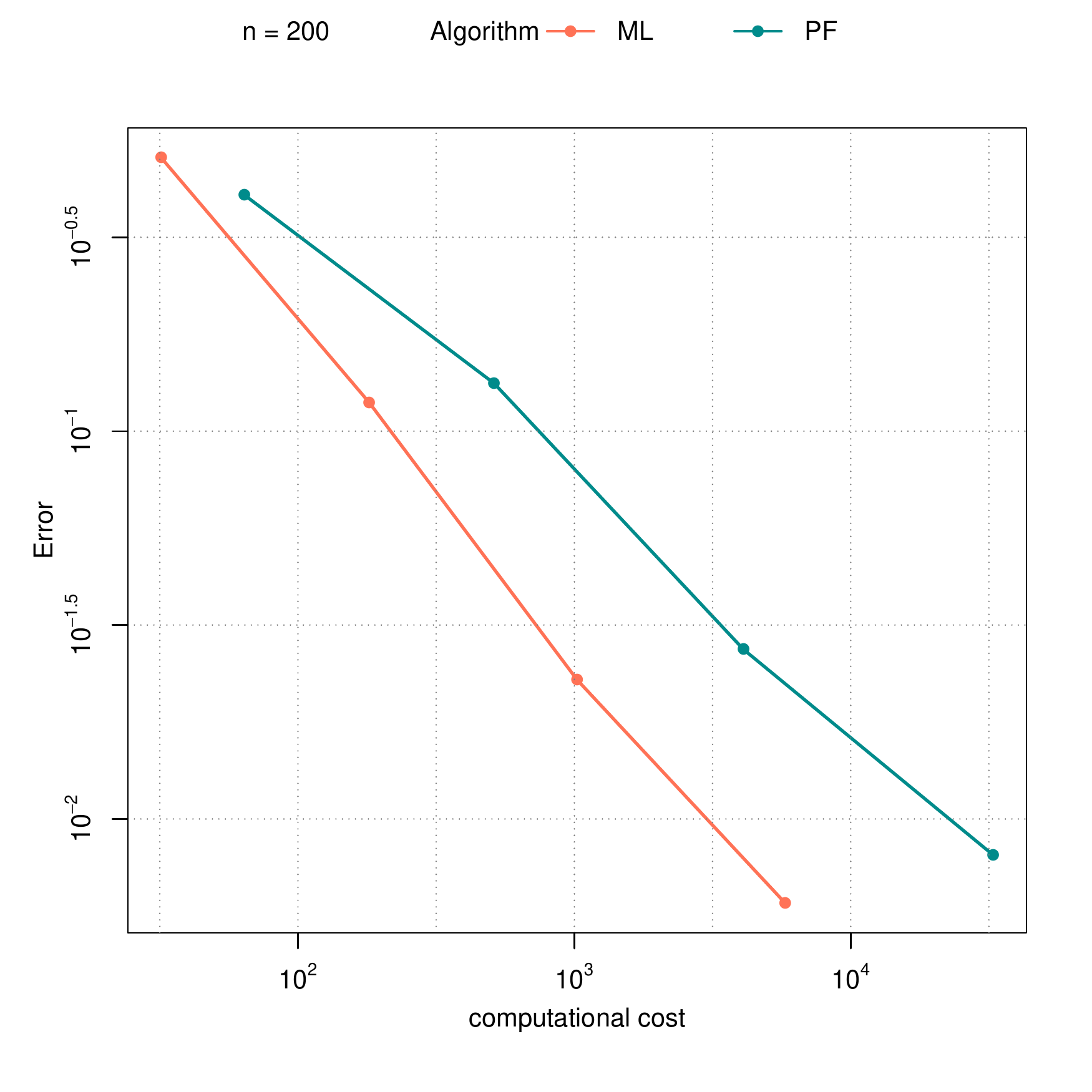}}}
	\caption{Computational savings for pricing barrier option with constant volatility using PF and MLPF; the constant volatility $\sigma=1.25$.}
	\label{ch3:fig2}
\end{figure}

The underlying process follows a geometric Brownian motion as above.  The performance of both PF and MLPF can be seen in Figure \ref{ch3:fig2}.  It can be observed that at all different time points considered, the MLPF achieves significant computational savings against the PF estimates.  This agrees with the theoretical conlcusions contained in \cite{Jay_mlpf,Jay_mlnormconst}. Note that these latter results do not consider the time parameter ($n$), but as seen here, the improvement seems to be uniform in time
as conjectured in that work.

\subsubsection{Stochastic Volatility Model}
The underlying asset price follows a system
\begin{align*}
\mathrm{d}S_t&=rS_t\mathrm{d}t+\sigma V_tS_t\mathrm{d}W_t\\
\mathrm{d}V_t&=\frac{1}{2}{\bf{\nabla}}\log\pi\left(V_t\right)\mathrm{d}t+\beta\mathrm{d}B_t,
\end{align*}
where $\sigma,\beta >0$ are scale parameters, $\pi\left(V_t\right)$ is the probability density chosen to be the Student t-distribution with degrees of freedom $\nu=100$.  $S_t$ is the price.

The following initial values were used in the simulation of Algorithms \ref{algo:ch3:SPF} and \ref{algo:ch3:ML}; $s_0=32,v_0=1.25$, the strike price $K=30$ and the scale parameters $\sigma=0.25,\beta=0.75$.  

The same settings for introducing the potential function in the case of constant volatility is adopted.  It provides stable weights with minimal or no resampling at all time points considered.  These weights guide particle into regions of interest and prevent the zero payoffs before the terminal time.  

\begin{figure}[!ht]
	\centering
	\subfigure{{\includegraphics[height=7cm]{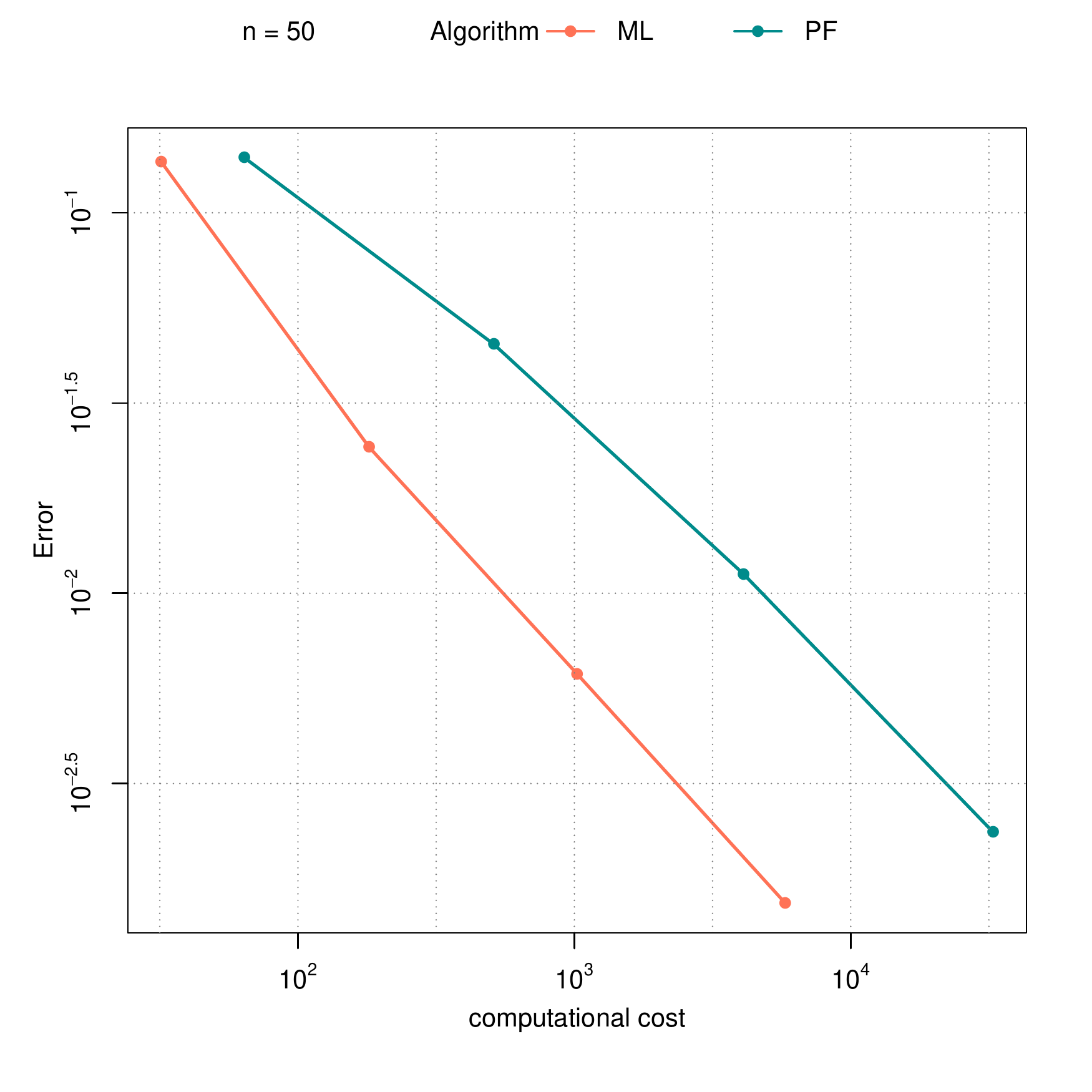}}}\,\,\,\,
	\subfigure{{\includegraphics[height=7cm]{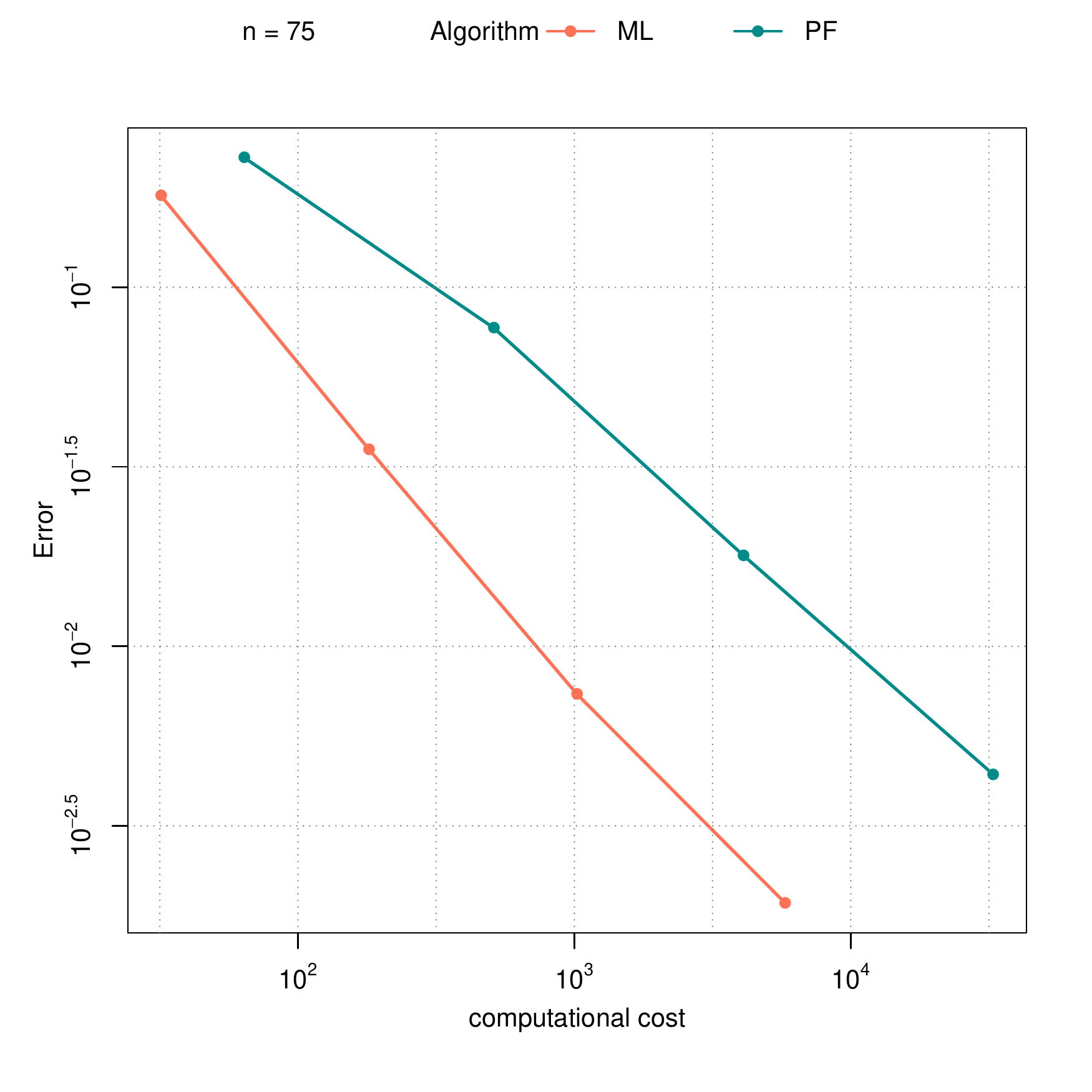}}}
	\subfigure{{\includegraphics[height=7cm]{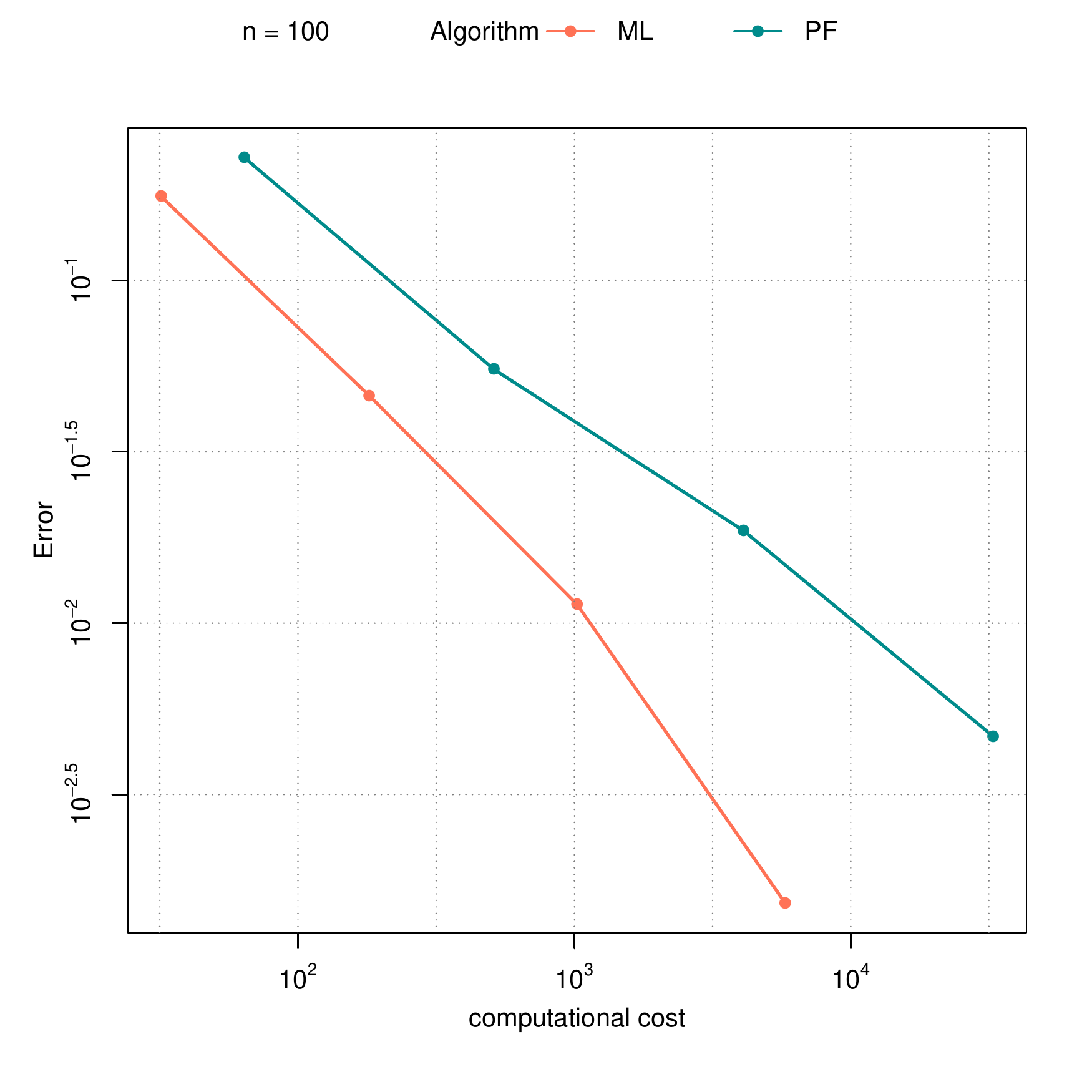}}}
	\subfigure{{\includegraphics[height=7cm]{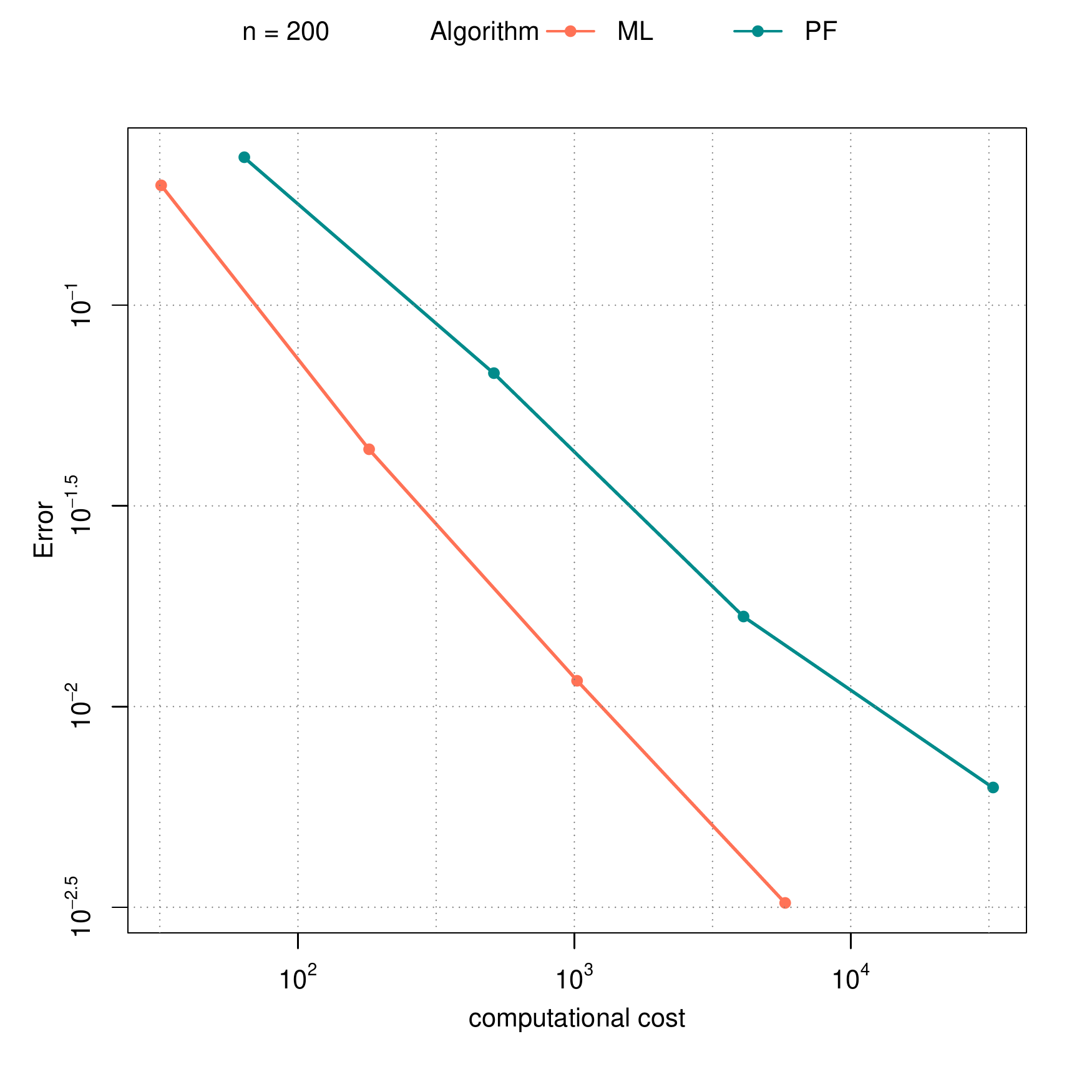}}}
	\caption{Computational savings for pricing barrier option with stochastic volatility using PF and MLPF; the SDE used is Langevin.}
	\label{ch3:fig3}
\end{figure}

In Figure \ref{ch3:fig3}, again, the MLPF outperfoms the PF at all different time units considered.  It is observed that as the time parameter increases, the error increases slightly but in general, the error rates are consistent with the theoretical findings in the multilevel set-up literature.

\subsection{TARNs}
We model the volatility as deterministic and stochastic differential equation.  The possible discontinuity of the payoff function of the TARNs is the main challenge in using the standard MC methods.  For illustration purposes, we consider a discontinous function $f$ of the form
\begin{align*}
f(s)&=\begin{cases}
2(s-60)+5 & \hbox{for}\thickspace s>60,\\
2(30-s)+5 & \hbox{for}\thickspace s<40,\\
-5 & \hbox{for}\thickspace 40\leqslant s\leqslant 60.
\end{cases}
\end{align*}
When standard MC is used in this case,  most of the samples stay inside the interval $\left(40,60\right)$, which could possibly leads to a zero payoff for the particle.  For example, the contribution for the first ten fixing dates is $-50$.  However, there is ocassional escapes of some of the particles within the first ten fixing dates and this contributes values significantly different from $-50$ due to discontinuity of the payoff function.  This makes the variance of the MC estimates very high.  We use $\tilde{g}=g$ and with an obvious truncation to $n$ variables.

\subsubsection{Constant Volatility}
The underlying asset price process follows the same stochastic differential equation used in the case of barrier options with constant volatility model.  

Several constant values of the volatility were used to check the performance of the estimated prices from the two algorithms.  We notice that the TARNs favored lower values of volatility compared to the barrier options, we do not show these results here for the same conclusions has been made independently in \cite{Sen_option}.  We are interested in the computational savings gained while using the multilevel set-up over the standard particle filter.

\begin{figure}[!ht]
	\centering
	\subfigure{{\includegraphics[height=7cm]{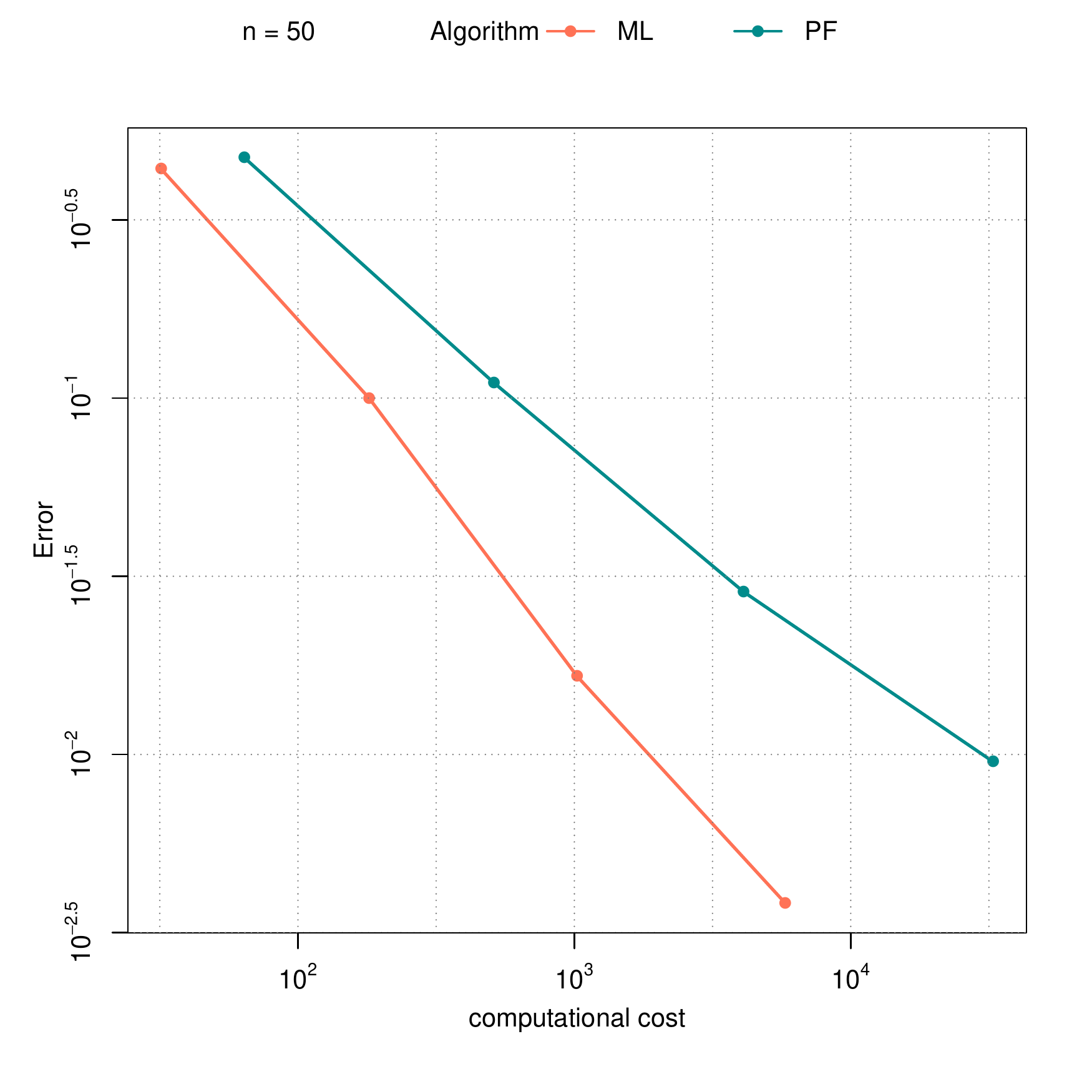}}}\,\,\,\,
	\subfigure{{\includegraphics[height=7cm]{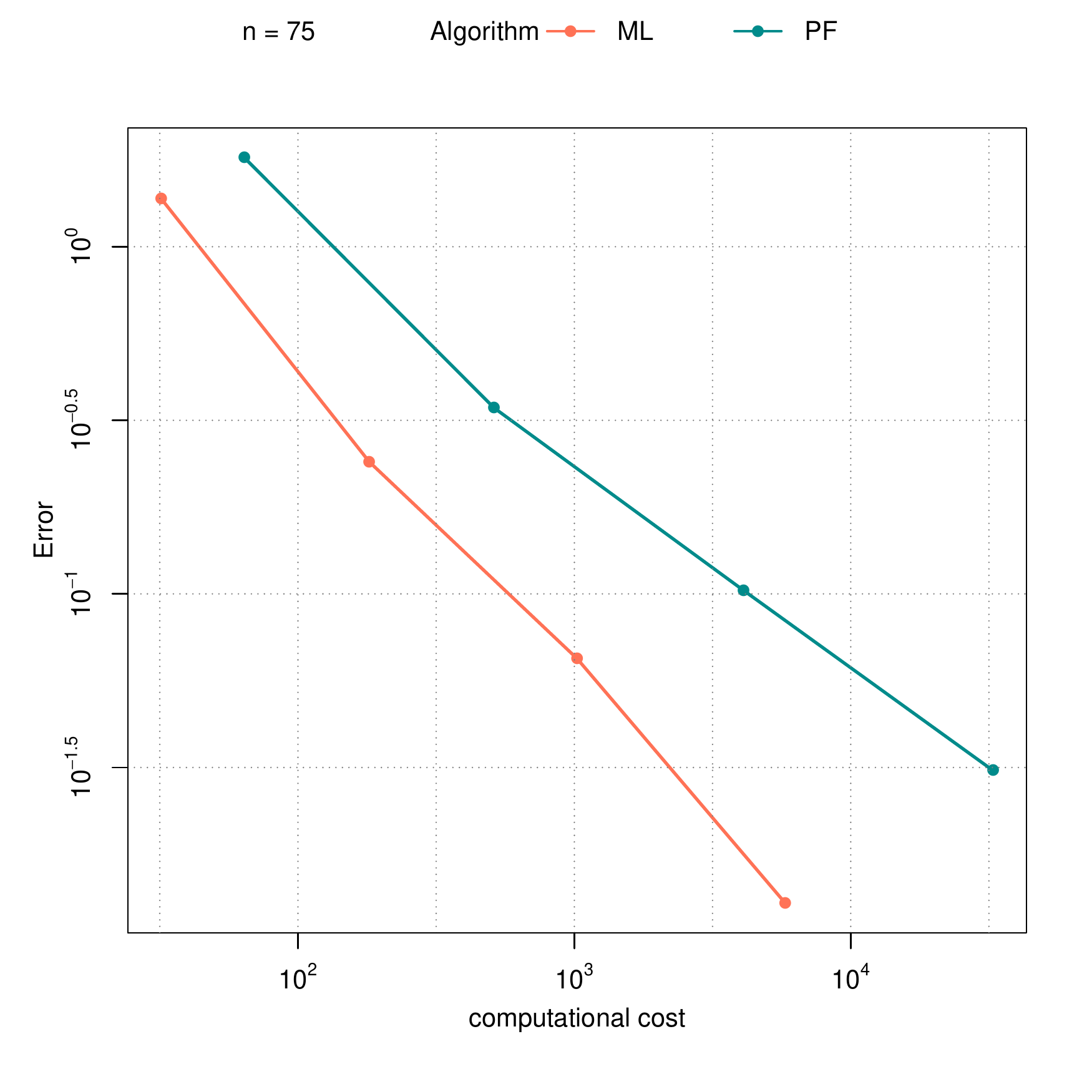}}}
	\subfigure{{\includegraphics[height=7cm]{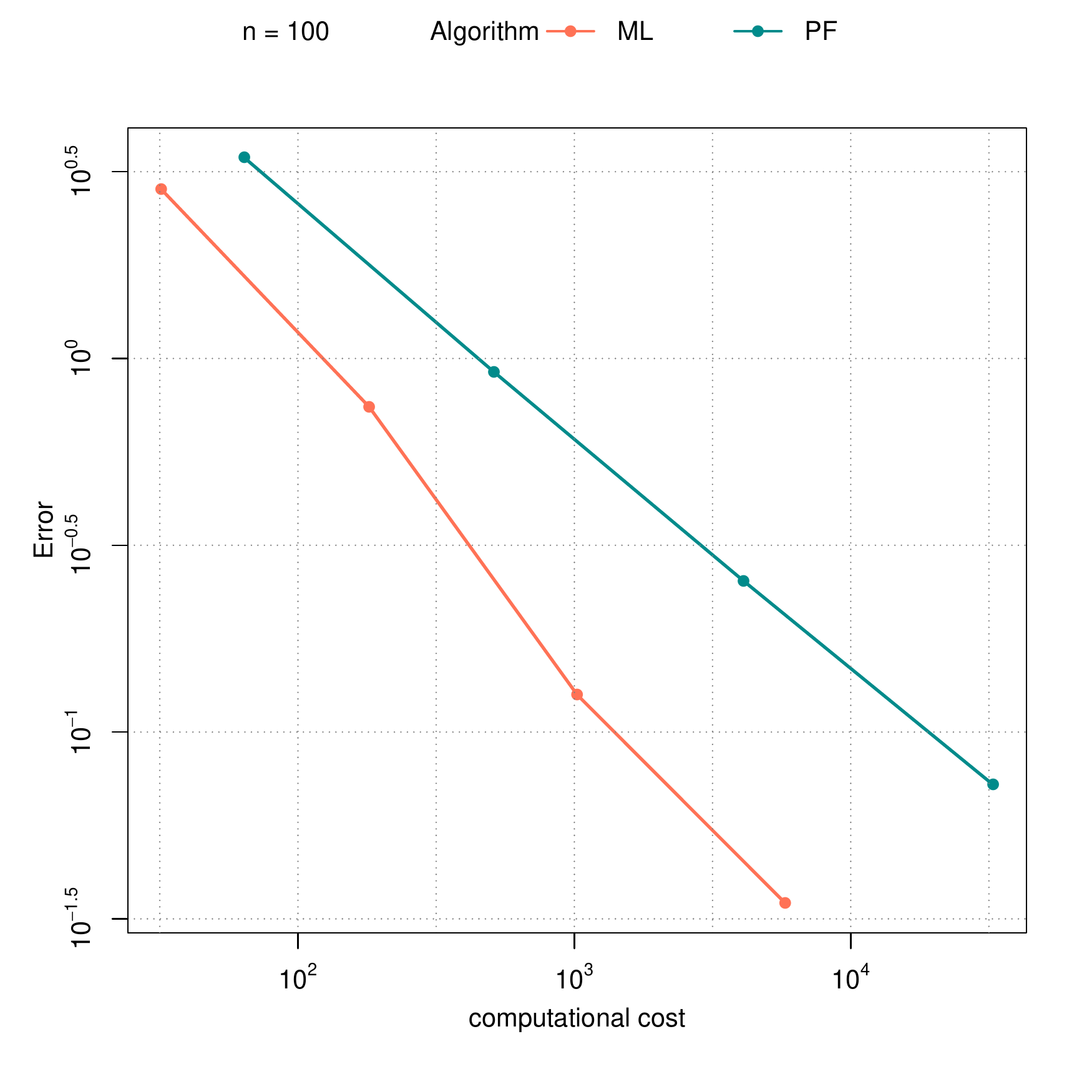}}}
	\subfigure{{\includegraphics[height=7cm]{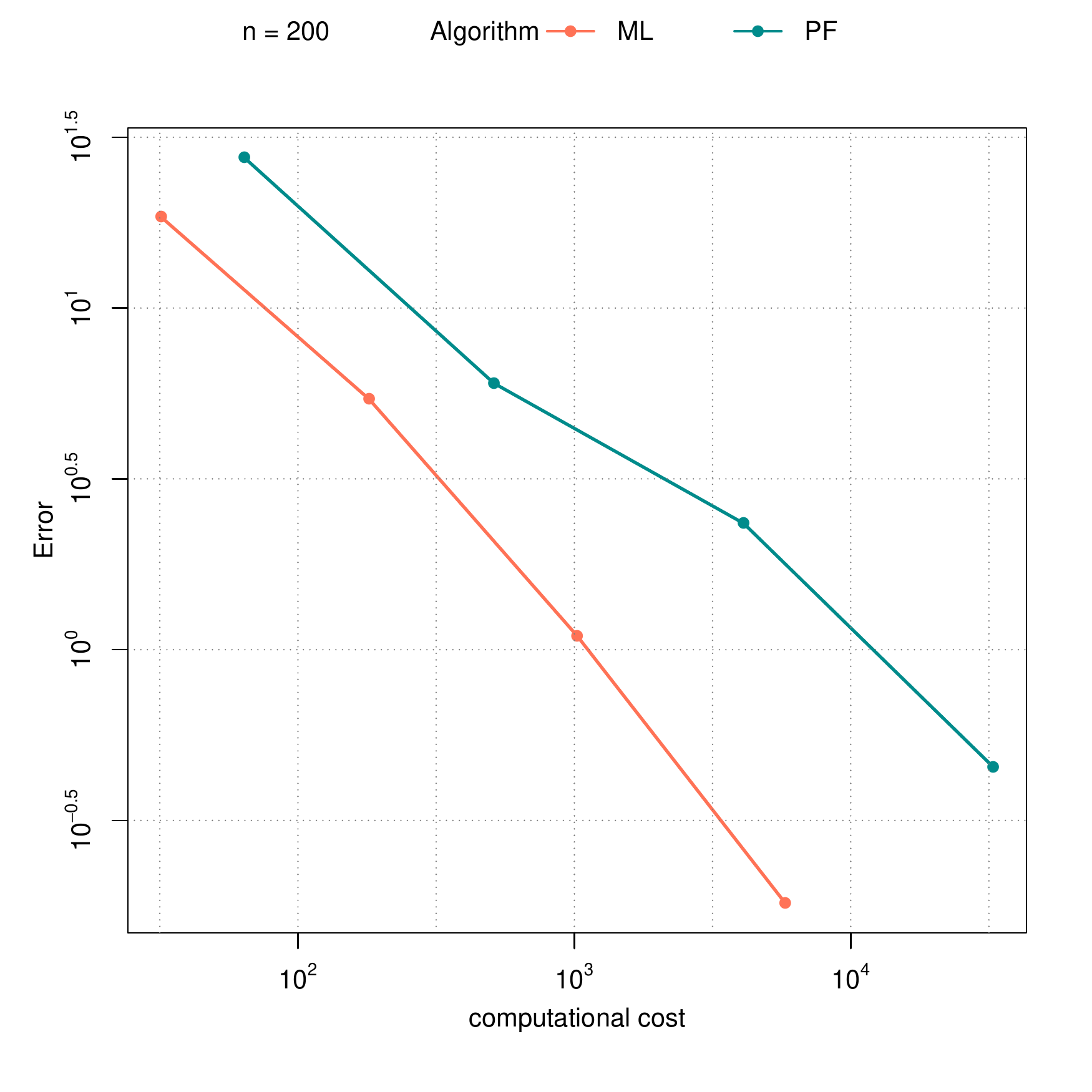}}}
	\caption{Computational cost plotted against MSE for TARNs with constant volatility.}
	\label{ch3:fig4}
\end{figure}

In Figure \ref{ch3:fig4}, the error plotted against the computational cost for both multilevel particle filter and standard particle filter can be seen.  The following initial values were adopted; $N_0=30, S_0=32$, where $N_0$ is  the notional value and a constant volatility of $\sigma=0.015625$.  The error can be seen as slighlty increasing from left to right as fixing dates increases.  It is clear from all the different time points considered that the MLPF has a significant advantage over PF in terms of computational savings.

\subsubsection{Stochastic Volatility Model}
The underlying financial asset follows the same system of stochastic differential equations used in the barrier option case with stochastic volatility model.  We use the same Langevin SDE for the underlying volatility for the TARNs.

\begin{figure}[!ht]
	\centering
	\subfigure{{\includegraphics[height=7cm]{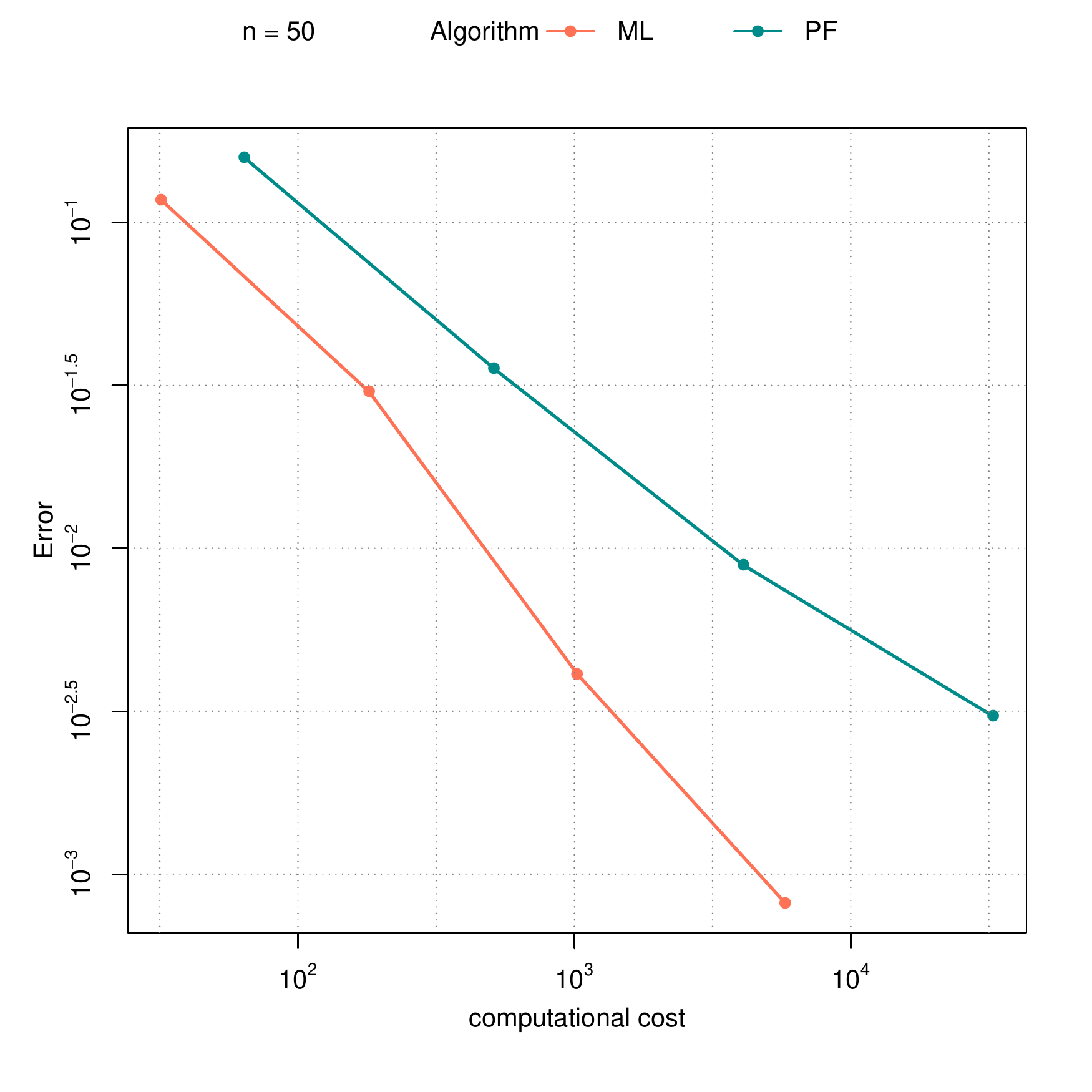}}}\,\,\,\,
	\subfigure{{\includegraphics[height=7cm]{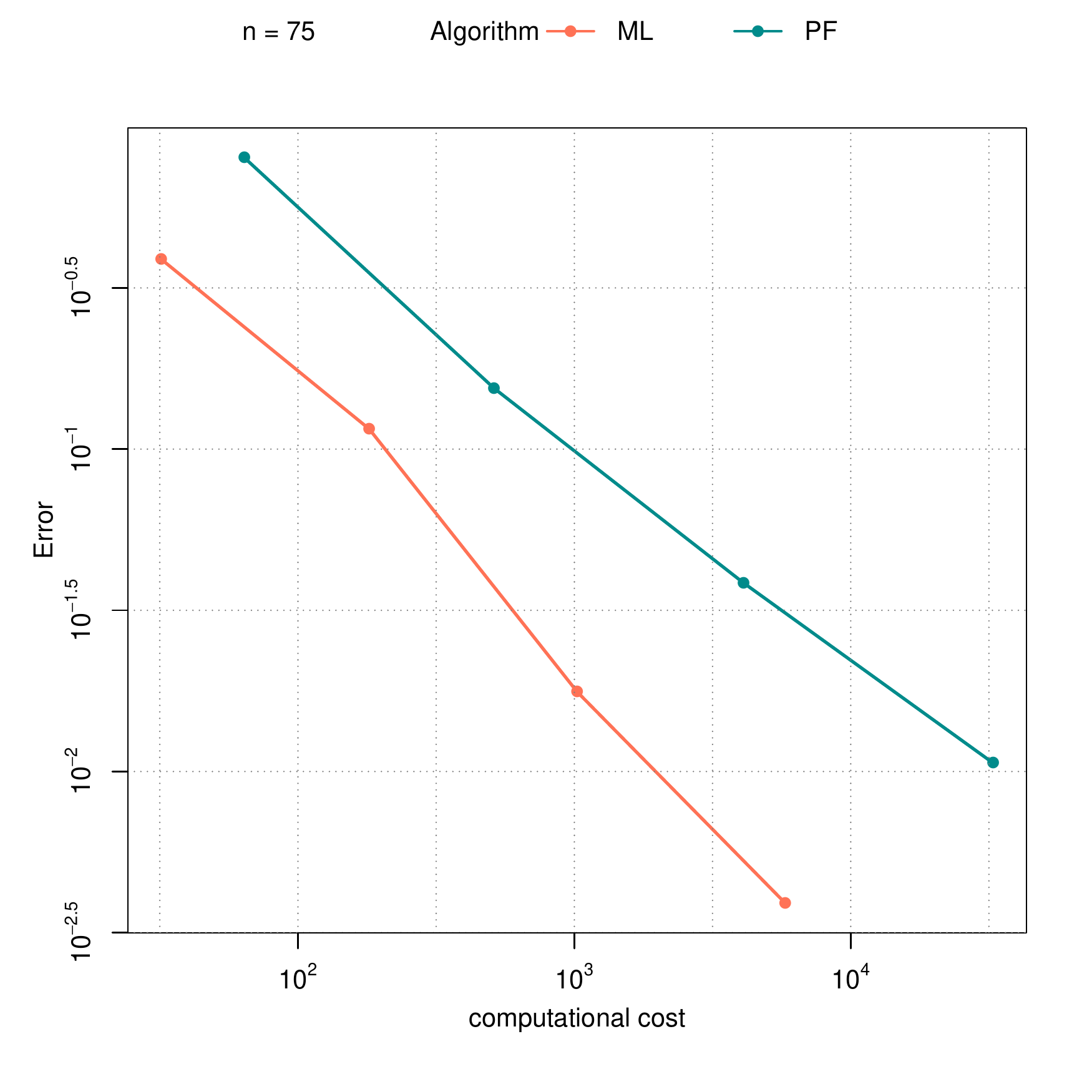}}}
	\subfigure{{\includegraphics[height=7cm]{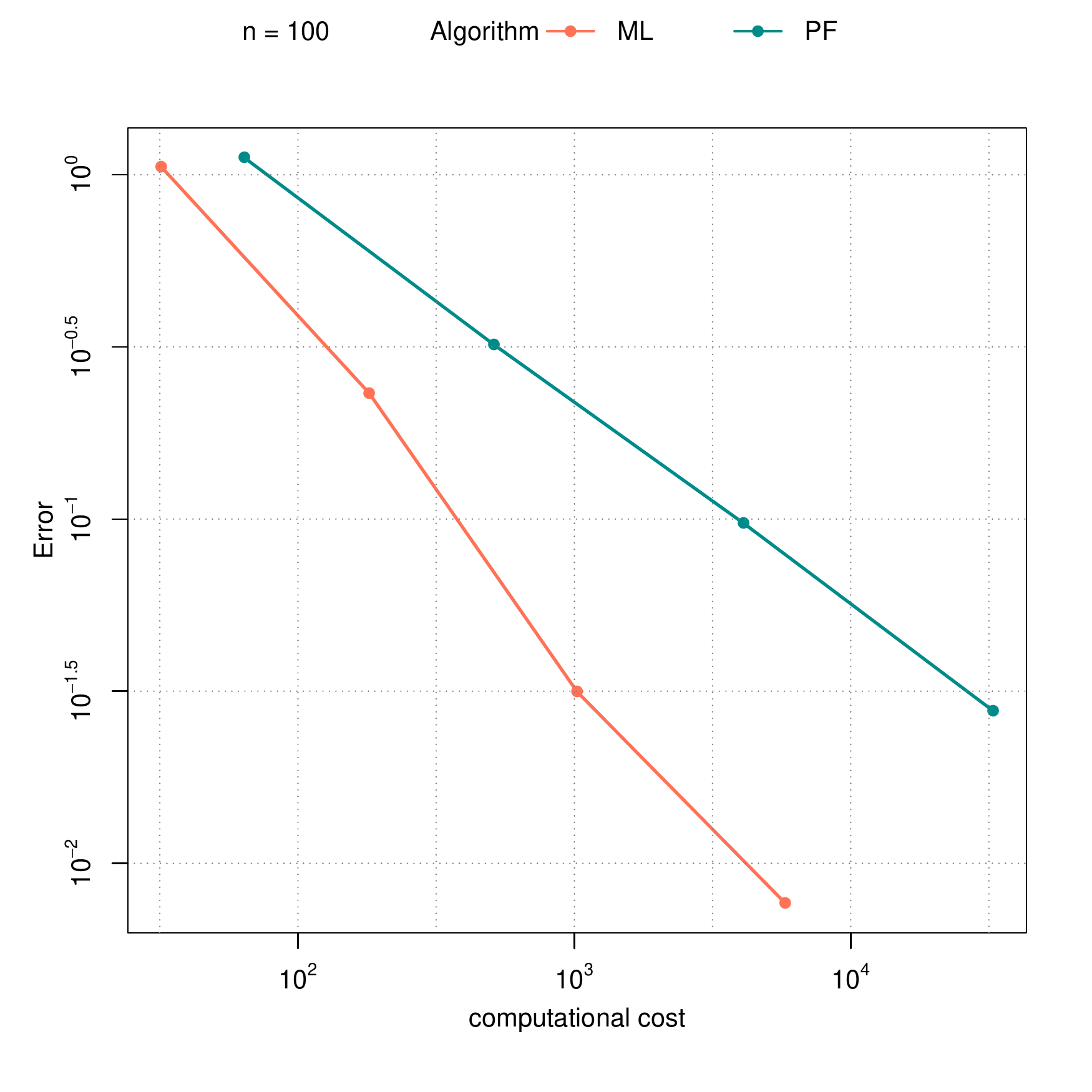}}}
	\subfigure{{\includegraphics[height=7cm]{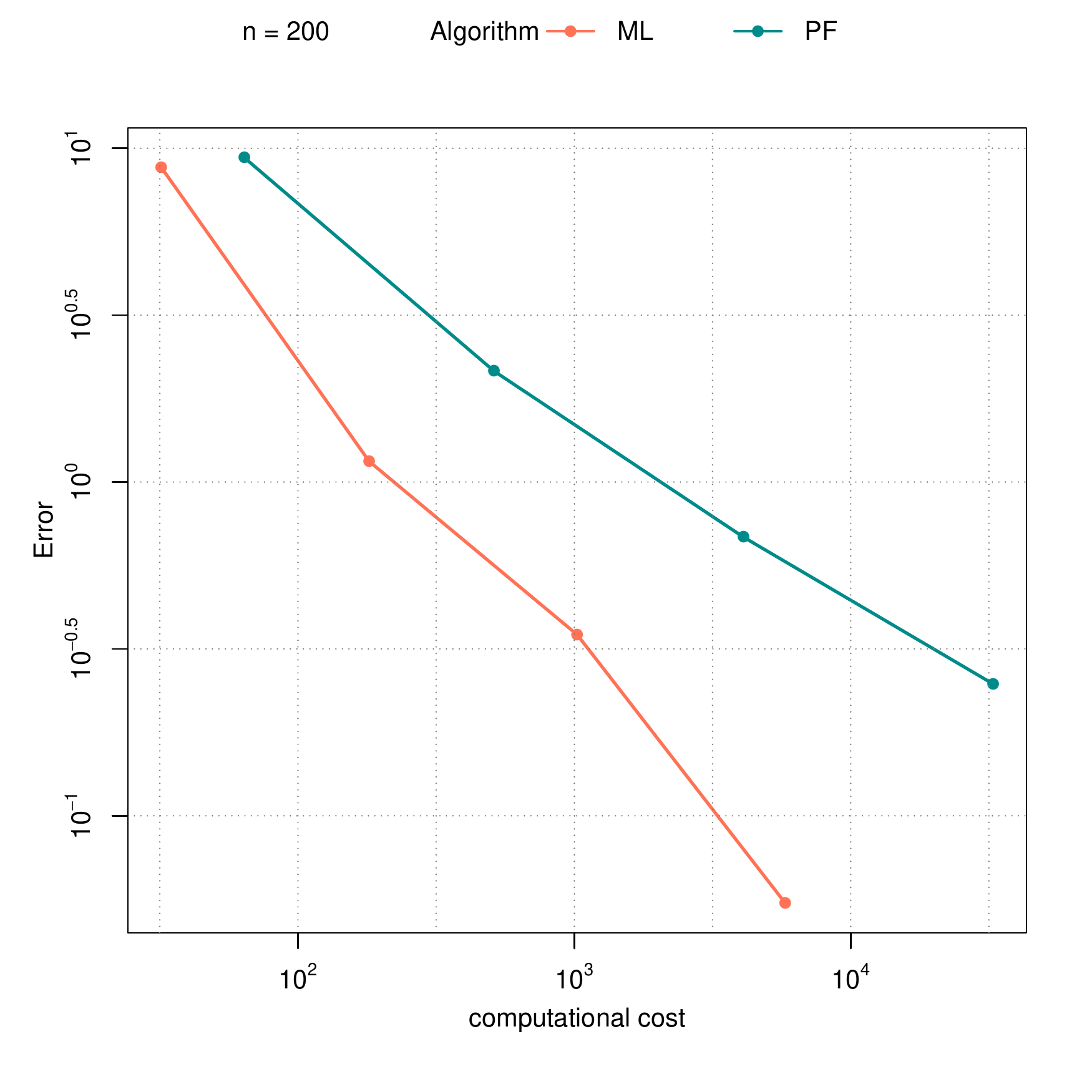}}}
	\caption{Computational savings gained in using multilevel particle filter against the standard particle filter with SDE volatility; TARNs with $50, 75,100$ and $200$ different times with a Notional value $N_0=30$.}
	\label{ch3:fig5}
\end{figure}

In Figure \ref{ch3:fig5}, the results of computational savings in using multilevel against standard partilce filter can be seen.

\section{Summary}\label{ch3:sec5}
In this paper, we have shown how the PF and MLPF can be used to estimate the price of both vanilla and exotic European type options.  We have demonstrated through several examples, the computational savings obtained when MLPF is used compared to the standard particle filter.

The methods presented here enhance the existing standard SMC methods when one seek to leverage in an optimal way in nested problems arising from multilevel set-up.  Throughout the article, one underlying financial asset was considered and it will be of interest to apply the MLPF on basket of underlying financial assets.  The ideas presented here can be extended to more complex stochastic volatility and interest rates processes.

\subsubsection*{Acknowledgements}
AJ was supported by Ministry of Education AcRF tier 2 grant, R-155-000-161-112. 

\appendix

\section{Sampling the Coupling}

Given the SDEs \eqref{ch3:eq1} we describe how to sample the Euler discretized coupling. We consider a pair of levels $(l,l-1)$, with $h_l=2^{-l}$ for any $l\geq 2$.
To sample the discretization at level $l$ up to some time $t$ we suppose this induces $k_l$ discretized points. Consider $W(m)\stackrel{i.i.d.}{\sim}\mathcal{N}(0,1)$ 
($\mathcal{N}(0,1)$ is the standard normal distribution) and independently $B(m)\stackrel{i.i.d.}{\sim}\mathcal{N}(0,1)$, $m\in\{0,\dots,k_{l}\}$. Then given the point
$(v_0,s_0)=(v_0^l,s_0^l)=(v_0^{l-1},s_0^{l-1})$ (note that one can easily have $(v_0^l,s_0^l)\neq(v_0^{l-1},s_0^{l-1})$ for the finer discretization, we have the recursion,
for $m\in\{0,\dots,k_l\}$
\begin{eqnarray*}
S_{m+1}^l & = & S_m^l + \alpha(S_m^l)h_l + \beta(S_m^l,V_m^l)\sqrt{h_l} W(m) \\
V_{m+1}^l & = & V_m^l + \gamma(V_m^l)h_l + \nu(V_m^l)\sqrt{h_l} B(m).
\end{eqnarray*}
For the more coarse trajectory, for $m\in\{0,\dots,k_{l-1}\}$
\begin{eqnarray*}
S_{m+1}^{l-1} & = & S_m^{l-1} + \alpha(S_m^{l-1})h_{l-1} + \beta(S_m^{l-1},V_m^{l-1})\sqrt{h_{l-1}}[W(2m)+W(2m+1)] \\
V_{m+1}^{l-1} & = & V_m^{l-1} + \gamma(V_m^{l-1})h_{l-1} + \nu(V_m^{l-1})\sqrt{h_{l-1}}[B(2m)+B(2m+1)].
\end{eqnarray*}


\begin{thebibliography}{9}

\bibitem{Barndorff_OUbased}
{\sc Barndorff-Nielsen}, O. E., \& {\sc Shephard}, N.~(2001). Non-Gaussian OU-based models and some of their uses in financial economics. \textit{J. R. Statist. Soc. B}, {\bf 63}, 167--241

\bibitem{Boyle}
{\sc Boyle}, P.~(1977). Options: A Monte Carlo approach. \textit{J. Fin. Econ.}, {\bf 4}, 323--338.

\bibitem{cdg:11}
{\sc Cerou}, F., {\sc Del Moral,} P., \& {\sc Guyader}, A.~(2011).
A non-asymptotic theorem for unnormalized Feynman-Kac particle models.
\emph{Ann. Inst. Henri Poincaire}, {\bf 47}, 629--649.

\bibitem{delm:04}
{\sc Del Moral}, P.~(2004). \textit{Feynman-Kac Formulae: Genealogical and
Interacting Particle Systems with Applications}. Springer: New York.

\bibitem{doucet_tutorial}
{\sc Doucet}, A. \& {\sc Johansen}, A. (2011). A tutorial on particle filtering and smoothing: Fifteen years later. In \emph{Handbook of Nonlinear Filtering} (eds. D. Crisan et B. Rozovsky), Oxford University Press: Oxford.


\bibitem{giles2008mlmc}
{\sc Giles}, M. B.~(2008). Multilevel Monte Carlo path simulation.  \textit{Oper. Res} {\bf 56}(3), 607--617.

\bibitem{Giles_mlmc}
{\sc Giles}, M. B.~(2015). Multilevel Monte Carlo methods. \textit{Acta Numerica}, {\bf 24}, 259--328.

\bibitem{glasserman_mcfinance}
{\sc Glasserman}, P.~(2003).
Monte Carlo Methods in Financial Engineering. Springer, New York.

\bibitem{Glasserman_barrieroptions}
{\sc Glasserman}, P., \& {\sc Staum}, J.~(2001). Conditioning on one-step survival for barrier options. \emph{Op. Res.}, {\bf 49}, 923--937.


\bibitem{JayOption}
{\sc Jasra}, A. \& {\sc Del Moral} P.~(2011). Sequential Monte Carlo methods for option pricing. \emph{Stoch. Anal. Appl.}, {\bf 29}, 292--316.


\bibitem{JayLevy}
{\sc Jasra}, A., {\sc Stephens}, D., {\sc Doucet}, A., \& {\sc Tsagaris}, T.~(2011).
Inference for L\'{e}vy driven stochastic volatility models via adaptive Sequential Monte Carlo. \emph{Scand. J. Statist.}, {\bf 38}, 1--22.

\bibitem{Jay_smcdiff}
{\sc Jasra}, A. \& {\sc Doucet} A.~(2009).
Sequential Monte Carlo methods for diffusion processes. \emph{Proc. Roy.} \emph{Soc. A}, {\bf 465}, 3709--3727.

\bibitem{Jay_mlpf}
{\sc Jasra}, A., {\sc Kamatani}, K., {\sc Law}, K. J. H., \& {\sc Zhou} Y.~(2015). Multilevel particle filters. \emph{arXiv:1510.04977}.

\bibitem{Jay_mlnormconst}
{\sc Jasra}, A., {\sc Kamatani}, K., {\sc Osei}, P. P., \& {\sc Zhou} Y.~(2017). Multilevel particle filters: Normalizing Constant Estimation.  \emph{Statist. Comp.} (to appear).


\bibitem{Sen_coupling}
{\sc Sen}, D., {\sc Thiery}, A., \& {\sc Jasra} A.~(2017).  On coupling particle filter trajectories. \emph{Statist. Comp.} (to appear).

\bibitem{Sen_option}
{\sc Sen}, D., {\sc Jasra}, A., \& {\sc Zhou} Y.~(2016). Some contributions to Sequential Monte Carlo methods for Option Pricing. \emph{J. Statist. Comp. Sim}, {\bf 87}, 733--752.

\bibitem{shev_op}
{\sc Shevchenko}, P., \& {\sc Del Moral}, P.~(2016).
Valuation of Barrier Options using Sequential Monte Carlo.
\emph{J. Comp. Fin.}, {\bf 20}, 1--29.


\end{thebibliography}
\end{document}